\documentclass[journal]{IEEEtran}
\usepackage{cite}
\usepackage[pdftex]{graphicx}
\usepackage{amsmath,bm}
\usepackage{amsthm}
\usepackage{amssymb}
\usepackage{algorithm}
\usepackage{algpseudocode}
\usepackage{array}
\usepackage{amsfonts}
\usepackage{optidef}
\usepackage{lineno,hyperref}
\usepackage{float}
\usepackage{tabularx}
\usepackage{xcolor}
\makeatletter
\let\MYcaption\@makecaption
\makeatother

\usepackage[font=footnotesize]{subcaption}

\makeatletter
\let\@makecaption\MYcaption
\makeatother

\usepackage{slashbox}
\usepackage{multirow}
\usepackage{color}

\newtheorem{theorem}{Theorem}
\newtheorem{lemma}{Lemma}

\begin{document}

\title{A Fast Iterative Method for Removing Impulsive Noise from Sparse Signals}

\author{Sahar~Sadrizadeh,
        Nematollah Zarmehi,
        Ehsan Asadi,
        Hamidreza Abin,
        and~Farokh~Marvasti
\thanks {Author are with the Advanced Communication Research
Institute (ACRI), Electrical Engineering Department, Sharif University of Technology,
Tehran, Iran (email: ss.sadrizadeh@ee.sharif.edu; zarmehi\_n@ee.sharif.edu; ea460@cam.ac.uk; hamidreza.abin@ee.sharif.edu; marvasti@sharif.edu). E. Asadi is currently pursuing PhD degree at the University of Cambridge}}

\markboth{Journal of \LaTeX\ Class Files,~Vol.~14, No.~8, August~2018}%
{S. Sadrizadeh: An Iterative Method for Removing Impulsive Noise from Sparse Signals}

\maketitle

\begin{abstract}
In this paper, we propose a new method to reconstruct a signal corrupted by noise where both signal and noise are sparse but in different domains.
The problem investigated in this paper arises in different applications such as impulsive noise removal
from images, audios and videos, decomposition of low-rank and sparse components of matrices, and separation of texts from images. First, we provide a cost function for our problem and then present an iterative method to find its local minimum.
The analysis of the algorithm is also provided. As an application of this problem, we apply our algorithm for impulsive noise Salt-and-Pepper noise (SPN) and Random-Valued Impulsive Noise (RVIN)) removal from images and compare our results with other notable algorithms in the literature. Furthermore, we apply our algorithm for removing clicks from audio signals. Simulation results show that our algorithms is simple and fast, and it outperforms other state-of-the-art methods in terms of reconstruction quality and/or complexity.
\end{abstract}


\section{Introduction} \label{introduction}
\IEEEPARstart{T}{he} problem considered in this paper can be modeled as:
\begin{equation}\label{eq:1}
  \mathbf{Y}={{\mathcal{D}}^{-1}}\left( {\mathbf{X_0}} \right) + {\mathbf{N_0}},
\end{equation}
where the original signal ${{\mathcal{D}}^{-1}}\left( {{\mathbf{X_0}}} \right)\in \mathbb{R}^{m\times n}$ is corrupted additively by sparse noise $\mathbf{N_0}\in \mathbb{R}^{m\times n}$, and $\mathcal{D}$ is the domain in which the signal is sparse; in other words, the signal and the noise are both sparse but in different domains.
We aim to reconstruct the original signal by removing the impulsive noise from the observed signal $\mathbf{Y}$.

One of the applications of this model is impulsive noise removal from images, videos and audios since these signals are sparse in some domains such as Discrete Cosine Transform (DCT), Wavelet and Contourlet; the impulsive noise is sparse in the image domain. Random missing samples, SPN noise and RVIN (common types of sparse noise) are common phenomenon in image processing,
audio and video transition \cite{zahedpour2009impulsive,hosseini2015real,toh2010noise}, and data transition over noisy communication channels \cite{kim2012blind,lyu2019robust} such as underwater acoustic channels and power line channels. Additionally, in dictionary learning problems where impulsive noise exists \cite{gao2018q,javaheri2018robust} (such as random missing samples when no side information about the location of missing samples is available) the problem can be modeled by equation (\ref{eq:1}).
The model stated in (\ref{eq:1}) also arises in low-rank and sparse matrix decomposition since the singular values of a low-rank matrices are sparse.\cite{zarmehi2018sparse,oymak2015simultaneously}. Another application of our model is separation of text from images since the text is sparse in the space domain while the image is sparse in the DCT domain.

The algorithms related to the impulsive noise removal can be divided into two general categories:

In the first category, methods first detect the position of the corrupted samples, i.e., the position of the impulsive noise, and then restore them from other clean samples. In the case of SPN, most of the research are from this category and usually result in a better reconstruction since they first find the mask matrix with which the signal is corrupted \cite{hosseini2015real,adler2012audio,roy2016impulse,gupta2015random,wu2011pde,ahmed2014removal,khaw2018high}. These methods have two drawbacks: When the original signal is corrupted by RVIN, the detection of noisy pixels becomes very challenging. In addition, all these methods utilize the structure of the audio and image signals (mainly their low-pass characteristic) to detect the location of corrupted pixels and hence they are not applicable to signals other than audio and image signals. As examples of this category, inpainting of audio signals corrupted by impulsive noise is considered in \cite{adler2012audio}. It is assumed that the location of the distorted data is known and the audio signal is reconstructed through sparse recovery techniques. In \cite{hosseini2015real}, noisy pixels are detected through an impulse detector and the image is restored by applying a weighted-average filter. The authors of \cite{roy2016impulse} present a two-step algorithm. In the first step the noisy pixels are detected by Support Vector Machine (SVM) classification, and then they are restored by applying an adaptive fuzzy filter.

The second category consists of methods which detect and restore
the noisy samples simultaneously \cite{zahedpour2009impulsive,hwang1995adaptive,avila2012bayesian,chou2013turbulent,wu2014random,
chen2016structure}. The method presented in this paper falls into this category and we compare our results with other algorithms of this class. Examples of this category are as follows: In \cite{hwang1995adaptive}, the Adaptive Median Filter (AMF) is introduced for impulsive noise removal from images. In this algorithm, the window size of the median filter is adjusted according to the impulsive noise
density. An Adaptive Median Filter which utilizes the Center-Weighted median (ACWMF) is introduced in \cite{chen2001adaptive}, and unlike other median-based filters, it performs well in the presence of RVIN. In \cite{jiang2014mixed}, the Weighted Encoding with Sparse Nonlocal Regularization  method (WESNR) is introduced which integrates a soft impulse detection and sparse non-local prior to remove mixed noise from images. The authors of \cite{avila2012bayesian} present a method based on Bayesian inference for impulsive noise removal from audio signals. For restoring images corrupted by impulsive noise, a method is suggested in \cite{chou2013turbulent} which utilizes particle swarm optimization and fuzzy filtering. The Structure-Adaptive Fuzzy Estimation (SAFE) algorithm is introduced in \cite{chen2016structure}, in which RVIN is removed via Gaussian Maximum Likelihood Estimation. The structure information of the image is incorporated into this algorithm as the fuzziness metrics in the form of point reliability and structure similarity. The Annihilating filter-based Low-Rank Hankel Matrix Approach (ALOHA) is proposed in \cite{jin2018sparse}. This method models the impulsive noise as a sparse component, and the underlying image is modeled as a low-rank Hankel structured matrix.

In this paper, a new iterative method is proposed which is applicable to 1-D and 2-D (even higher dimensions) sparse signals. Our algorithm can reconstruct signals which are corrupted by any type of sparse noise such as SPN and RVIN, by contrast to most of the other methods which are applicable to one of these noises. Moreover, the noisy samples are not detecting beforehand. Our method reconstructs both signal and noise iteratively by thresholding them in their corresponding sparse domains and projecting them onto the set imposed by (\ref{eq:1}). The contributions of this paper are summarized as follows:

1) In this paper, we propose a general framework for impulsive noise removal. The only assumption about the noise and signal is their sparsity in different domains. Therefore, the proposed algorithm can easily be applied to various applications (any dimension) with different noise models. To the best of our knowledge, this is one of the few works with this generality.

2) Since the proposed model \eqref{eq:1} is non-convex, we presented an iterative algorithm called Iterative Double Thresholding (IDT) to approximate the solution. The capability of our algorithm is verified analytically and experimentally.

3) By proposing some modifications, a fast method is obtained for impulsive noise removal from images and audios.

The rest of this paper is organized as follows. In Section \ref{method}, first, a cost function is introduced that generates our algorithm through optimization and then it is analysed. The simulation results are discussed in Section \ref{simulation} with comparisons to other methods in the literature. Finally, the paper is concluded in Section \ref{conclusion}.


\section{The Proposed Method} \label{method}

In this section, we illustrate the proposed scheme for removing sparse noise from sparse signal. Then we analyse our algorithm and present some modifications to tailor our applications.

\subsection{Iterative Double Thresholding Algorithm} \label{algorithm}

As mentioned in the introduction, we are considering the problem of reconstructing a signal which is sparse in some domain from its noisy observation; the noise is also assumed to be sparse in the observation domain. As an example of \eqref{eq:1}, the transformation $\mathcal{D}$, the original signal and the noise can be the 2-dimenional DCT, an image and impulsive noise, respectively. Therefore, the problem becomes impulsive noise removal from images.

We can consider this problem from another view: we have an overcomplete dictionary (by concatenating matrix representation of the ${\mathcal{D}}$ transform and the identity matrix), and our goal is to find the sparsest representation of the matrix $\mathbf{Y}$ \cite{donoho2001uncertainty}.

For an observed matrix $\mathbf{Y}$, there may exist infinite numbers of 2-tuples $(\mathbf{X},\mathbf{N})$ that satisfy \eqref{eq:1}, but we are looking for the sparsest pair, that is, the pair with the minimum total number of non-zero entries. This specific 2-tuple is the minimizer of the following optimization problem:
\begin{equation}\label{eq:27}
\begin{aligned}
& \underset{{(\mathbf{X},\mathbf{N})}}{\text{argmin}}
& & \|\text{vec}(\mathbf{X})\|_0 + \|\text{vec}(\mathbf{N})\|_0, \\
& \quad \text{s.t.}
& & (\mathbf{X},\mathbf{N})\in\mathrm{W},
\end{aligned}
\end{equation}
where $\|\text{vec}(.)\|_0$ represents the $L_0$ semi-norm of the vectorization of the input matrix, and the set $\mathrm{W}$ contains all the $(\mathbf{X},\mathbf{N})$ for which \eqref{eq:1} holds:
\begin{equation}\label{eq:3}
  \mathrm{W}\triangleq\{(\mathbf{X},\mathbf{N})|\mathbf{Y}={{\mathcal{D}}^{-1}}\left( {\mathbf{X}} \right) + {\mathbf{N}}\};
\end{equation}
If for the parameters of our problem, the solution of \eqref{eq:1} is unique, we can easily conclude that the desired pair ${(\mathbf{X_0}},\mathbf{N_0})$ is the sparsest member of $\mathrm{W}$. We will discuss the uniqueness of the solution later in this section.

This optimization problem \eqref{eq:27} is non-convex and NP-hard; hence we present an alternative optimization problem for finding the sparsest member of $\mathrm{W}$. Consider the following function:

\begin{equation}\label{eq:4}
\begin{split}
   {f_{{\lambda}}}( {{\mathbf{X}},{\mathbf{N}},{\mathbf{T}_{\mathbf{1}}},{\mathbf{T}_{\mathbf{2}}}} )&
   \triangleq\|(\mathbf{1}-{\mathbf{T}_{\mathbf{1}}})\odot\mathbf{X}\|_F^{2}+\|(\mathbf{1}-{\mathbf{T}_{\mathbf{2}}})\odot\mathbf{N}\|_F^{2}\\
     &+\lambda(\|\text{vec}({\mathbf{T}_{\mathbf{1}}})\|_1 + \|\text{vec}({\mathbf{T}_{\mathbf{2}}})\|_1),
\end{split}
\end{equation}
where $(\mathbf{X},\mathbf{N})\in\mathrm{W}$, ${\mathbf{T}_{\mathbf{1}}},{\mathbf{T}_{\mathbf{2}}}$ are binary matrices, and $\mathbf{1}$ is a  matrix of all ones. The sign $\odot$ represents the Hadamard (entry-wise) product of matrices, and $\|\text{vec}(.)\|_1$ denotes the entry-wise $L_1$ norm (or the $L _1$ norm of the vectorization) of the input matrix.
We will prove later in this section that for small enough values of $\lambda$, the minimizer of the following optimization problem is the sparsest member of $\mathrm{W}$, i.e., the solution of \eqref{eq:27}:
\begin{equation}\label{eq:5}
\begin{aligned}
& \underset{{(\mathbf{X},\mathbf{N}),({\mathbf{T}_{\mathbf{1}}},{\mathbf{T}_{\mathbf{2}}})}}{\text{argmin}}
& & {{f_{{\lambda}}}( {{\mathbf{X}},{\mathbf{N}},{\mathbf{T}_{\mathbf{1}}},{\mathbf{T}_{\mathbf{2}}}} )}. \\
& \quad \text{s.t.}
& & (\mathbf{X},\mathbf{N})\in\mathrm{W}.
\end{aligned}
\end{equation}
In the first two terms of the cost function, $\mathbf{1}-{\mathbf{T}_{\mathbf{1}}},\mathbf{1}-{\mathbf{T}_{\mathbf{2}}}$ equal the complement of the binary matrices ${\mathbf{T}_{\mathbf{1}}},{\mathbf{T}_{\mathbf{2}}}$, respectively, and thus these two terms compute the sum of squares of the elements of $(\mathbf{X},\mathbf{N})$ which are outside the supports ${\mathbf{T}_{\mathbf{1}}},{\mathbf{T}_{\mathbf{2}}}$. The last two terms of the cost function calculate the total number of non-zero entries of ${\mathbf{T}_{\mathbf{1}}},{\mathbf{T}_{\mathbf{2}}}$. Therefore, we are looking for two binary matrices  ${\mathbf{T}_{\mathbf{1}}},{\mathbf{T}_{\mathbf{2}}}$ with minimum total number of non-zero entries which are closest to the support (location of non-zero entries) of the two matrices $(\mathbf{X},\mathbf{N})\in \mathrm{W}$. The parameter $\lambda$ balances the weight of the two parts of the cost function. It is worth mentioning that this function is convex \emph{w.r.t} $(\mathbf{X},\mathbf{N})$ when $({\mathbf{T}_{\mathbf{1}}},{\mathbf{T}_{\mathbf{2}}})$ are fixed since the feasible region is a convex set, and the objective function is a quadratic function of $(\mathbf{X},\mathbf{N})$.

Now we present our algorithm for finding the minimizer of \eqref{eq:5}. The pseudocode of the IDT algorithm is illustrated in Algorithm \ref{alg:1}.
\begin{algorithm}[!tb]
\caption{IDT}
\label{alg:1}
\begin{algorithmic}[1]
\State \textbf{Input}:
\State \quad Observed matrix: $\mathbf{Y}\in \mathbb{R}^{m\times n}$
\State \quad Maximum number of iterations of the outer loop: $K$
\State \quad Decreasing sequence: $\mathbf{th} \in \mathbb{R}^{K}$
\State \quad Stopping threshold of the inner loop: $\delta$
\State \textbf{Output}:
\State \quad Recovered estimate of the signal: $\mathbf{\hat{X}}$
\State \quad Recovered estimate of the noise: $\mathbf{\hat{N}}$
\Procedure{}{}
\State $\mathbf{\hat{X}}\leftarrow {\mathcal{D}}({\mathbf{Y}})$, \quad ${{\mathbf{\hat{N}}}} \leftarrow \mathbf{0}$
\For {$k = 1 : K$}
\State $\mathbf{X}^{0}\leftarrow {{\mathbf{\hat{X}}}}$, \quad ${{\mathbf{N}}^{0}} \leftarrow {{\mathbf{\hat{N}}}}$, \quad $l\leftarrow 0$
\State $\sqrt{\lambda}\leftarrow \mathbf{th}_k$
\While {$e > \delta$}
\State ${{\mathbf{X}}^{l+1}}\leftarrow$ threshold($ {|{{\mathbf{X}}^l_{i,j}}|}^{m,n}_{i=1,j=1} , \sqrt{\lambda}$)
\State ${{\mathbf{N}}^{l+1}}\leftarrow$ threshold($ {|{{\mathbf{N}}^l_{i,j}}|}^{m,n}_{i=1,j=1} , \sqrt{\lambda}$)
\State ${{\mathbf{X}}^{l + 1}} \leftarrow 0.5\;\left( {{{\mathbf{X}}^{l+1}} + {\cal D}\left( {{\mathbf{Y}} - {{\mathbf{N}}^{l+1}}} \right)} \right)$
\State ${{\mathbf{N}}^{l + 1}} \leftarrow 0.5\;\left( { - {{\cal D}^{ - 1}}\left( {{{\mathbf{X}}^{l+1}}} \right) + {\mathbf{Y}} + {{\mathbf{N}}^{l+1}}} \right)$
\State $e \leftarrow \|{{\mathbf{N}}^{l + 1}} - {{\mathbf{N}}^{l}}\|_F$
\State $l \leftarrow l + 1$
\EndWhile
\State ${{\mathbf{\hat{X}}}}\leftarrow {{\mathbf{X}}^{l}}$
\State ${{\mathbf{\hat{N}}}}\leftarrow {{\mathbf{N}}^{l}}$
\EndFor
\State\Return ${\mathbf{\hat X}}$ , ${\mathbf{\hat N}}$
\EndProcedure
\end{algorithmic}
\end{algorithm}

As it will be proved in the next subsection, we are looking for the minimizer of $f_{\lambda}$ for small enough value of $\lambda$. However, the cost function has many local minimums in this case. Therefore, making use of warm start is necessary. We start with a large $\lambda$ and iteratively minimize the cost function; then this estimation will be used as an initial guess for the next optimization with lower $\lambda$. It is worth mentioning that as $\lambda$ goes to infinity, the optimization problem becomes convex since the last two terms of the cost function are forced to become zero, and thus $({\mathbf{T}_{\mathbf{1}}},{\mathbf{T}_{\mathbf{2}}})$ are all zero matrices. Therefore, our optimization problem reduces to the following one, which is convex:
\begin{equation}\label{eq:28}
\begin{aligned}
& \underset{{(\mathbf{X},\mathbf{N})}}{\text{argmin}}
& & \|\mathbf{X}\|_F^{2}+\|\mathbf{N}\|_F^{2}. \\
& \quad \text{s.t.}
& & (\mathbf{X},\mathbf{N})\in\mathrm{W}.
\end{aligned}
\end{equation}

We initialize the algorithm for a large value of $\lambda$ in line 10. For a specific value of $\lambda$, in order to minimize $f_{\lambda}$, we will alternatively minimize the cost function $w.r.t$ $({\mathbf{T}_{\mathbf{1}}},{\mathbf{T}_{\mathbf{2}}})$ and $(\mathbf{X},\mathbf{N})$, then we project the resultant minimizers onto the set $\mathrm{W}$ so that the constraint of the optimization problem is satisfied.

For a fixed value of $\lambda$ (in the inner loop of the algorithm), let $\mathbf{X}^{l}$ and $\mathbf{N}^{l}$ denote the recovered signal and noise after $l$ iterations. In the $({\mathbf{T}_{\mathbf{1}}},{\mathbf{T}_{\mathbf{2}}})$-minimization step, the minimizers can be found by thresholding $(\mathbf{X},\mathbf{N})$ of the previous iteration:
\begin{equation}\label{eq:6}
\begin{split}
& ({\mathbf{\hat{T}}}_\mathbf{1})_{i,j} = \left\{ {\begin{array}{*{20}{c}}
{0,\;{(\mathbf{X}^{l})_{i,j}} < \sqrt {\lambda} }\\
{1,\;{(\mathbf{X}^{l})_{i,j}} \ge \sqrt {\lambda} }
\end{array}} \right.,\\ &({\mathbf{\hat{T}}}_\mathbf{2})_{i,j} = \left\{ {\begin{array}{*{20}{c}}
{0,\;{(\mathbf{N}^{l})_{i,j}} < \sqrt {\lambda} }\\
{1,\;{(\mathbf{N}^{l})_{i,j}} \ge \sqrt {\lambda} }
\end{array}} \right.,
\end{split}
\end{equation}
where $(.)_{i,j}$ is the element of the matrix at the intersection of the $i$-th row and the $j$-th column.

In the $({\mathbf{X}},{\mathbf{N}})$-minimization step, a new pair of $\mathbf{X}^{l+1}$ and $\mathbf{N}^{l+1}$ (which is not necessarily in $\mathrm{W}$) is attained by considering $({\mathbf{\hat{T}}_{\mathbf{1}}},{\mathbf{\hat{T}}_{\mathbf{2}}})$ to be their respective supports as follows:
\begin{equation}\label{eq:12}
\mathbf{X}^{l+1} = {\mathbf{\hat{T}}_{\mathbf{1}}} \odot \mathbf{X}^{l+1} \qquad , \qquad \mathbf{N}^{l+1} = {\mathbf{\hat{T}}_{\mathbf{2}}} \odot \mathbf{N}^{l+1}.
\end{equation}

The insertion of (\ref{eq:6}) in (\ref{eq:12}) is equivalent to thresholding the entries of $\mathbf{X}^{l}$ and $\mathbf{N}^{l}$ by $\sqrt {\lambda}$. This is done in the lines 15 and 16 of Algorithm \ref{alg:1}, where threshold(${|{{\mathbf{X}}_{i,j}}|}^{m,n}_{i=1,j=1} , th$) represents thresholding the entries of matrix $\mathbf{X}$ based on their absolute values with regard to the threshold level of $th$.

Now we have to project this pair onto the set $\mathrm{W}$ so that the constraint of the optimization problem (\ref{eq:5}) is met. We will use the following lemma and find the projection of $(\mathbf{X}^{l+1},\mathbf{N}^{l+1})$ on the set $\mathrm{W}$.

\begin{lemma}\label{lm:3}
The projection of an arbitrary pair $(\mathbf{X},\mathbf{N})$ onto the set $\mathrm{W}$ defined by \eqref{eq:3} is
\begin{equation}\label{eq:9}
\left\{ {\begin{array}{l}
{\hat{\mathbf{X}} = 0.5\;\left( {{\mathbf{X}}  + \mathcal{D}\left( {{\mathbf{Y}} - {\mathbf{N}}} \right)} \right)}\\
{\hat{\mathbf{N}} = 0.5\;( - {{\cal D}^{ - 1}}\left( {\mathbf{X}} \right) + \mathbf{Y} + \mathbf{N})}
\end{array}} \right..
\end{equation}
\end{lemma}
\begin{IEEEproof}
See Appendix A.
\end{IEEEproof}
The process of projection (\ref{eq:9}) is carried out in lines 17 and 18 of Algorithm \ref{alg:1}.

The procedure explained above is continued iteratively until the difference between two consecutive estimation $\|{{\mathbf{N}}^{l + 1}} - {{\mathbf{N}}^{l}}\|_F$ is less than a threshold. Then $\lambda$ is decreased (outer loop of the algorithm) and the minimizer of the new $f_\lambda$ is found in the same manner.

From another point of view, our algorithm gradually picks up the main components of the signal and noise. By decreasing parameter $\lambda$, i.e., decreasing the threshold level, we are increasing the support size of the estimated signal and noise.

\subsection{Algorithm Analysis}
As discussed at the beginning of this section, the proposed algorithm works if we could prove:

1) The solution of the optimization problems \eqref{eq:27} and \eqref{eq:5} are the same.

2) Under certain conditions, the sparse solution of \eqref{eq:1} is unique.

We will investigate these two problems in Theorem \ref{th:1} and Theorem \ref{th:2}, respectively.

We define $\varepsilon$ as follows: For a fixed pair of binary matrices $({\mathbf{T}_{\mathbf{1}}},{\mathbf{T}_{\mathbf{2}}})$, the minimizers of $f_{\lambda}$ are denoted by $( {\mathbf{X^*}},{\mathbf{N^*}})$ (which are computed in the algorithm by thresholding and projecting). Since the total number of binary matrices $({\mathbf{T}_{\mathbf{1}}},{\mathbf{T}_{\mathbf{2}}})$ is finite, there exists only finite numbers  of 4-tuples $({{\mathbf{X^*}}^i},{{\mathbf{N^*}}^i},{\mathbf{T}^{i}_{\mathbf{1}}},{\mathbf{T}^{i}_{\mathbf{2}}} )$, which are distinguished by superscript $i$. We are looking for the 4-tuple for which $f_{\lambda}$ is minimum. The minimum non-zero value of $\|(\mathbf{1}-{\mathbf{T}_{\mathbf{1}}})\odot\mathbf{X^*}\|_F^{2} +\|(\mathbf{1}-{\mathbf{T}_{\mathbf{2}}})\odot\mathbf{N^*}\|_F^{2} $ (first part of the cost function) among these 4-tuples is denoted by $\varepsilon$. Mathematically, $\varepsilon$ is defined as:
\begin{equation}\label{eq:13}
\begin{aligned}
\varepsilon \triangleq & \underset{{i}}{\text{ min}} \quad {\|(\mathbf{1}-{\mathbf{T}^i_{\mathbf{1}}})\odot{\mathbf{X^*}}^i\|_F^{2} + \|(\mathbf{1}-{\mathbf{T}^i_{\mathbf{2}}})\odot{\mathbf{N^*}}^i\|_F^{2}}. \\
& \text{ s.t.} \quad \|(\mathbf{1}-{\mathbf{T}^i_{\mathbf{1}}})\odot{\mathbf{X^*}}^i\|_F^{2} + \|(\mathbf{1}-{\mathbf{T}^i_{\mathbf{2}}})\odot{\mathbf{N^*}}^i\|_F^{2} \neq 0.
%
\end{aligned}\\
\end{equation}
\begin{theorem}\label{th:1}
Let the unique sparsest element of $\mathrm{W}$ (solution of \eqref{eq:27}) be $(\tilde{\mathbf{X}},\tilde{\mathbf{N}})$ with respective supports $({\tilde{\mathbf{T}}_{\mathbf{1}}},{\tilde{\mathbf{T}}_{\mathbf{2}}})$; moreover, let the sparsity numbers of $(\tilde{\mathbf{X}},\tilde{\mathbf{N}})$ be $(\tilde{k_1}, \tilde{k_2})$, then for $\lambda<\varepsilon/(\tilde{k_1}+\tilde{k_2})$, the minimizer of ${f_{{\lambda}}}
( {{\mathbf{X}},{\mathbf{N}},{\mathbf{T}_{\mathbf{1}}},{\mathbf{T}_{\mathbf{2}}}} )$ over $\mathrm{W}$ is $(\tilde{\mathbf{X}},\tilde{\mathbf{N}}, {\tilde{\mathbf{T}}_{\mathbf{1}}},
{\tilde{\mathbf{T}}_{\mathbf{2}}})$.
\end{theorem}

\begin{IEEEproof}
See Appendix B.
\end{IEEEproof}

Now we discuss the uniqueness of the sparsest member of $\mathrm{W}$ in Theorem \ref{th:2}.

\begin{theorem}\label{th:2}
The sparsest member of $\mathrm{W}$ with sparsity numbers ($\tilde{k_1},\tilde{k_2}$) is unique if the following sufficient conditions are satisfied:
\begin{equation}\label{eq:26}
\left\{ {\begin{array}{*{20}{c}}
{\mathbf{B}^\mathrm{T}\otimes \mathbf{A}} \text{ \emph{is orthonormal}} \\
{\tilde{k_1}+\tilde{k_2}< \frac{1}{2}(1+{\|\text{\emph{vec}}(\mathbf{B}^\mathrm{T}\otimes \mathbf{A})\|_\infty}^{-1}) }
\end{array}} \right.;
\end{equation}
where $(.)^\mathrm{T}$ and $\otimes$ denote the transpose of a matrix and the Kronecker product of two matrices, respectively; the linear transformation ${\mathcal{D}}^{-1}\left( \mathbf{X} \right)=\mathbf{A\,X\,B}$; and $\|\text{\emph{vec}}(.)\|_\infty$ represents the $L_\infty$ norm of the vectorization of the input, i.e., the entry with maximum absolute value.
\end{theorem}

\begin{IEEEproof}
See Appendix C.
\end{IEEEproof}

When ${\mathcal{D}}^{-1}$ is 2-dimensional inverse DCT (for images), we can easily conclude from Theorem \ref{th:2} that ${\tilde{k_1}+\tilde{k_2}< \frac{1}{2}(1+{\|\text{\emph{vec}}(\mathbf{D}\otimes \mathbf{D})\|_\infty}^{-1}) }$ guarantees the uniqueness of the sparsest member of $\mathrm{W}$, where $\mathbf{D}$ is the DCT-matrix. In the case of square matrices, $\mathbf{D} \in \mathbb{R}^{m\times m}$, we have:
\begin{equation}\label{eq:30}
\begin{split}
& \|\text{\emph{vec}}(\mathbf{D}\otimes \mathbf{D})\|_\infty = (\|\text{\emph{vec}}(\mathbf{D})\|_\infty)^2 =\\
&       \left\{ {\begin{array}{*{20}{lr}}
        \frac{2}{m}*cos^2(\frac{\pi}{2m}) \quad\text{m is power of 2} \\
        \quad\frac{2}{m} \qquad\qquad\qquad\text{o.w}
       \end{array}}\right.  ;
\end{split}
\end{equation}
Simulation results show that for much higher sparsity numbers, i.e., denser matrices, our algorithm still works.

\subsection{Modifications}\label{modification}
In this subsection, we proceed to introduce some modifications to our algorithm. The modified algorithm can be found in Algorithm \ref{alg:2}. Note that the lines of the algorithm marked with (*) are only for image denoising and should be omitted in other applications. It is seen through simulation results that these modifications make the algorithm faster and improve the reconstruction quality.

\begin{algorithm}[!tb]
\caption{Modified IDT}
\label{alg:2}
\begin{algorithmic}[1]
\State \textbf{Input}:
\State \quad Observed matrix: $\mathbf{Y}\in \mathbb{R}^{m\times n}$
\State \quad Four constants: $\alpha_1,\beta_1,\alpha_2,\beta_2$
\State \quad Standard deviation of the gaussian filter: $\sigma$ \hfill\((*)\)
\State \quad Maximum number of iterations: $K$
\State \quad Stopping threshold: $\delta$
\State \textbf{Output}:
\State \quad Recovered estimate of the signal: $\mathbf{\hat{X}}\in \mathbb{R}^{m\times n}$
\State \quad Recovered estimate of the noise: $\mathbf{\hat{N}}\in \mathbb{R}^{m\times n}$
\Procedure{}{}
\State $\mathbf{X}^0\leftarrow {\mathcal{D}}({\mathbf{Y}})$, \quad ${{\mathbf{N}}^0} \leftarrow 0$, \quad $k\leftarrow0$
\While {$e > \delta \;\& \;k \le K$}
\State ${{\mathbf{X}}^k}\leftarrow$ threshold($ {|{{\mathbf{X}}^k_{i,j}}|}^{m,n}_{i=1,j=1} , {\beta _1}{e^{ - {\alpha _1}k}}$)
\State ${{\mathbf{X}}^k} \leftarrow $ clip(${{\cal D}^{ - 1}}\left( {{{\mathbf{X}}^k}}\right)$) \hfill\((*)\)
\State ${{\mathbf{X}}^k} \leftarrow $ gaussian-filter(${{\mathbf{X}}^k}$ , $\sigma$) \hfill\((*)\)
\State ${{\mathbf{N}}^{k + 1}} \leftarrow {\mathbf{Y}} - {{{\mathbf{X}}^k}}$
\State ${{\mathbf{N}}^{k + 1}}\leftarrow$ threshold($ {|{{\mathbf{N}}^{k + 1}_{i,j}}|}^{m,n}_{i=1,j=1} , {\beta _2}{e^{ - {\alpha _2}k}}$)
\State ${{\mathbf{X}}^{k + 1}} \leftarrow {\cal D}\left( {{\mathbf{Y}} - {{\mathbf{N}}^{k + 1}}} \right)$
\State $e \leftarrow \|{{\mathbf{N}}^{k + 1}} - {{\mathbf{N}}^{k}}\|_F$
\State $k \leftarrow k + 1$
\EndWhile
\State\Return ${\mathbf{\hat X}} \leftarrow {{\mathbf{X}}^{k}}$, ${\mathbf{\hat N}} \leftarrow {{\mathbf{N}}^{k}}$
\EndProcedure
\end{algorithmic}
\end{algorithm}

In this algorithm, the inner loop and outer loop are merged. Furthermore, in the first algorithm, the signal and noise are first thresholded and then they are projected onto the set $\mathrm{W}$; however, in the modified version, the signal is first thresholded and an approximation of the sparse noise is derived according to (\ref{eq:1}), then this approximated noise is thresholded and a better estimation of the signal is found by (\ref{eq:1}). It should also be noted that two different threshold levels, $\mathbf{th1}$ and $\mathbf{th2}$, are considered for signal and noise. $\mathbf{th1}$ and $\mathbf{th2}$ can be any decreasing sequences, but as is common in other papers \cite{azghani2015microwave,zamani2016iterative,sadrizadeh2018iterative}, we adopted the exponential scheme to decrease them in each iteration. 

In the image processing context, the reconstruction is not perfect since images do not become purely sparse by DCT or other transforms used to make the images sparse such as Wavelet and Contourlet. Nevertheless, two pieces of side information are available, i.e., the signal values are in the interval $[0,255]$ for 8-bit images and the signal contains a large low frequency component. We can take advantage of the first one and clip the estimated signal in each iteration. Since the thresholding function is non-linear and changes the range of the signal and noise, this modification will result in a better reconstruction. In other words, we are considering the matrices with entries in $[0,255]$ and project our approximated signal onto this convex set in each iteration. Moreover, we can apply  a low-pass filter, to the signal so as to emphasize the low-pass component of the image and attenuate the high frequency components of the noise. Various filters were tested such as gaussian filter, median filter, adaptive median filter, non-local means \cite{buades2005non} and guided image filter \cite{he2013guided}. Gaussian filter was selected due to its best performance in our algorithm. It is worth noting that the filtering step can also be used in audio reconstruction.

\section{Simulation Results}\label{simulation}
\begin{figure}
  \centering
  \includegraphics[width=\linewidth]{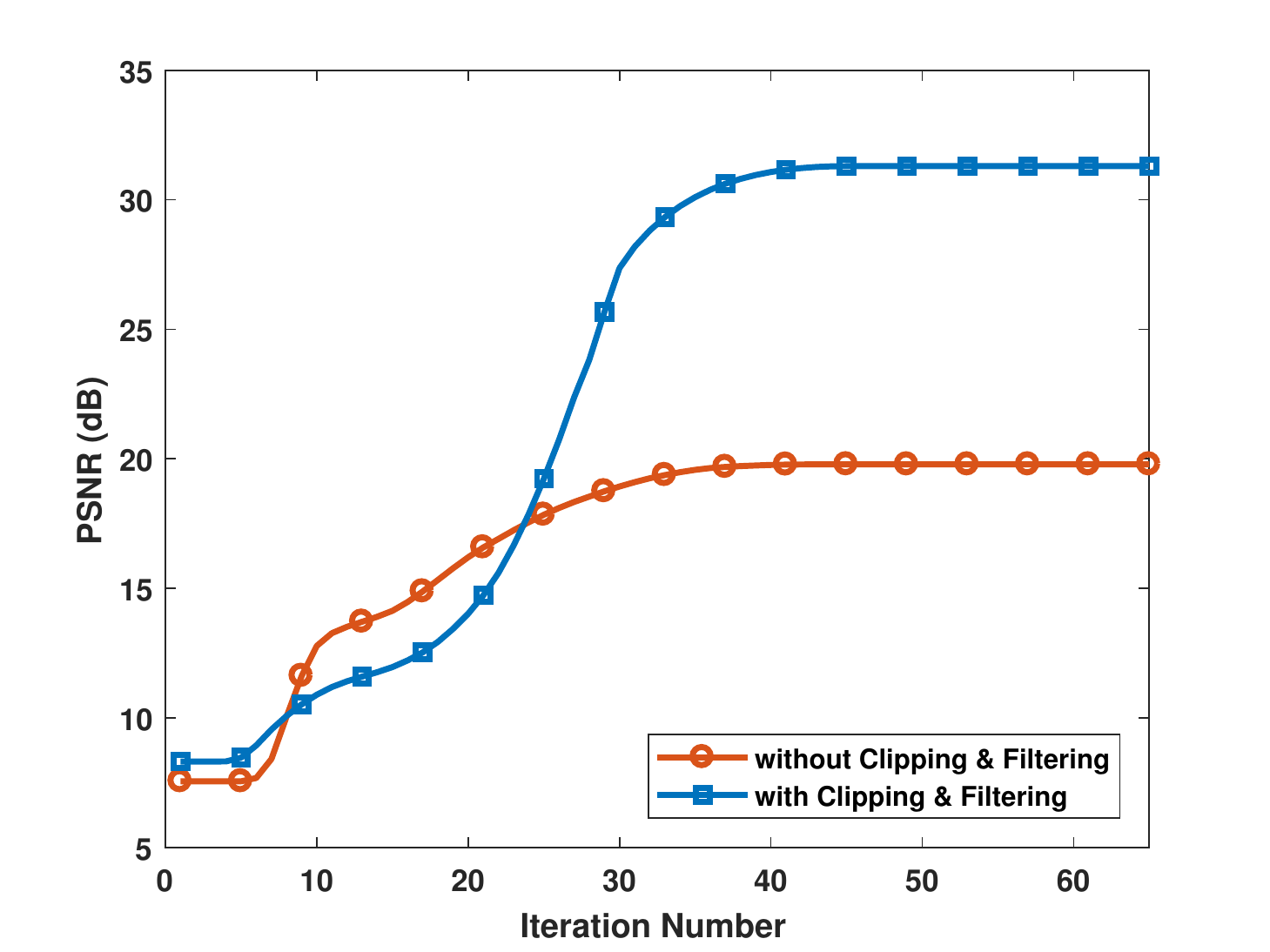}
  \caption{Effect of clipping and filtering for the case of 50\% salt-and-pepper noise for the Cropped Lena image.}\label{fig:2}
\end{figure}



We evaluate the effectiveness of our algorithm for different scenarios, i.e., images, audio and artificial sparse signals. Simulations are conducted in MATLAB 2018a on a PC equipped with an Intel Core i-7 3.60–GHZ CPU and 64-GB RAM.

\subsection{Parameter Setting}
We have 6 parameters in our algorithm: parameters of thresholding: $\alpha_1,\beta_1,\alpha_2,\beta_2$, standard deviation of the Gaussian filter: $\sigma$ and number of iterations: $K$. The thresholding parameters depend on the absolute values of the signal and noise in their respective sparse domain. As is common in impulsive noise removal
literature, in order to obtain a coarse estimation of the signal and noise, the adaptive median filters AMF \cite{hwang1995adaptive} and ACWMF \cite{chen2001adaptive} are applied to the observed noisy signal in the case of SPN and RVIN, respectively. The absolute value of the samples of the estimated signal and noise (in their sparse domain) with maximum absolute values are good choices for $\beta_1$ and $\beta_2$. Moreover, If we sort the samples of the estimated signal and noise based on their absolute values, the averages of the difference between two consecutive samples are good choices for $\alpha_1$ and $\alpha_2$.
We have come to this setting based on our experience: Our algorithm finds the larger components of the sparse signal in the first few iterations and by decreasing the threshold level, the lower components are found. Therefore, the proposed process of defining the parameters of thresholding seem reasonable and it works well in practice.
As the standard deviation of the Gaussian filter increases, the lower frequencies of the image are affected (the smoothing effect is higher). Hence, when the impulsive noise density is low, we use $\sigma = 0.4$ and in the case of higher noise densities, we use $\sigma = 0.55$. Finally, if the thresholding parameters are defined in a good manner, after at most $K=60$ iterations our algorithm converges. As $K$ signifies the run-time of the algorithm, we would like to minimize this parameter.

\subsection{Artificial Sparse Signal and Noise}\label{sparse}
In this subsection, we generate a $500\times 500$ 2-D signal which is sparse in DCT domain and a $500\times 500$ sparse noise. The sparse elements of the signal and noise are independently sampled from a normal distribution with variance 128. We evaluate the performance of our algorithm in terms of Success Rate (SR) and SNR. A reconstruction is considered as a success if the output SNR is greater than 60 dB. The results are presented in Table \ref{tab:1} for different signal and noise sparsity ratios $\rho = \mathrm{Sparsity\ Number}/500^2$. As can be seen in this table, when the sparsity ratio of both signal and noise are 30\%, the success rate falls despite the fact that the average SNR is still high. In order to explain this phenomenon, let the sparsity ratio of one of the signals be 30\%. Since the location and the value of these entries are unknown, there are about 60\% unknown variables and at least 60\% equations are needed to find the unknown variables. If the location of the non-zero entries of the other signal was known, its sparsity ratio could have been 40\%. However, in our case where no side information about the sparsity of the signals is provided, the sparsity ratio should be less than 40\%.
\begin{table}
\renewcommand{\arraystretch}{1.1}
\caption{Reconstruction SNR  ({\textrm{\normalfont dB}}) and SR (in the bracket) for Different Signal ($\rho_x$) and Noise Sparsity Ratios ($\rho_n$)}
\label{tab:1}
\centering
\begin{tabular}{|c||c|c|c|}
    \hline

    \backslashbox{$\rho_x$}{$\rho_n$}  & $\mathbf{10\%}$ & $\mathbf{20\%}$ & $\mathbf{30\%}$\\
    \hline
    \hline
    $\mathbf{10\%}$ & 316.5 (100\%) & 313.5 (100\%) & 311.6 (100\%) \\
    \hline
    $\mathbf{20\%}$ & 315.9 (100\%) & 312.6 (100\%) & 310.4 (100\%) \\
    \hline
    $\mathbf{30\%}$ & 314.9 (100\%) & 311.4 (100\%) & 224.082 (73\%) \\
    \hline

\end{tabular}
\end{table}
Table \ref{tab:1} shows that when the signal and noise are purely sparse, our algorithm can fully reconstruct the signal.

\subsection{Impulsive Noise Removal from Images}\label{Images}
\begin{figure*}
    \centering 
\begin{subfigure}{0.3\textwidth}
  \includegraphics[width=\linewidth]{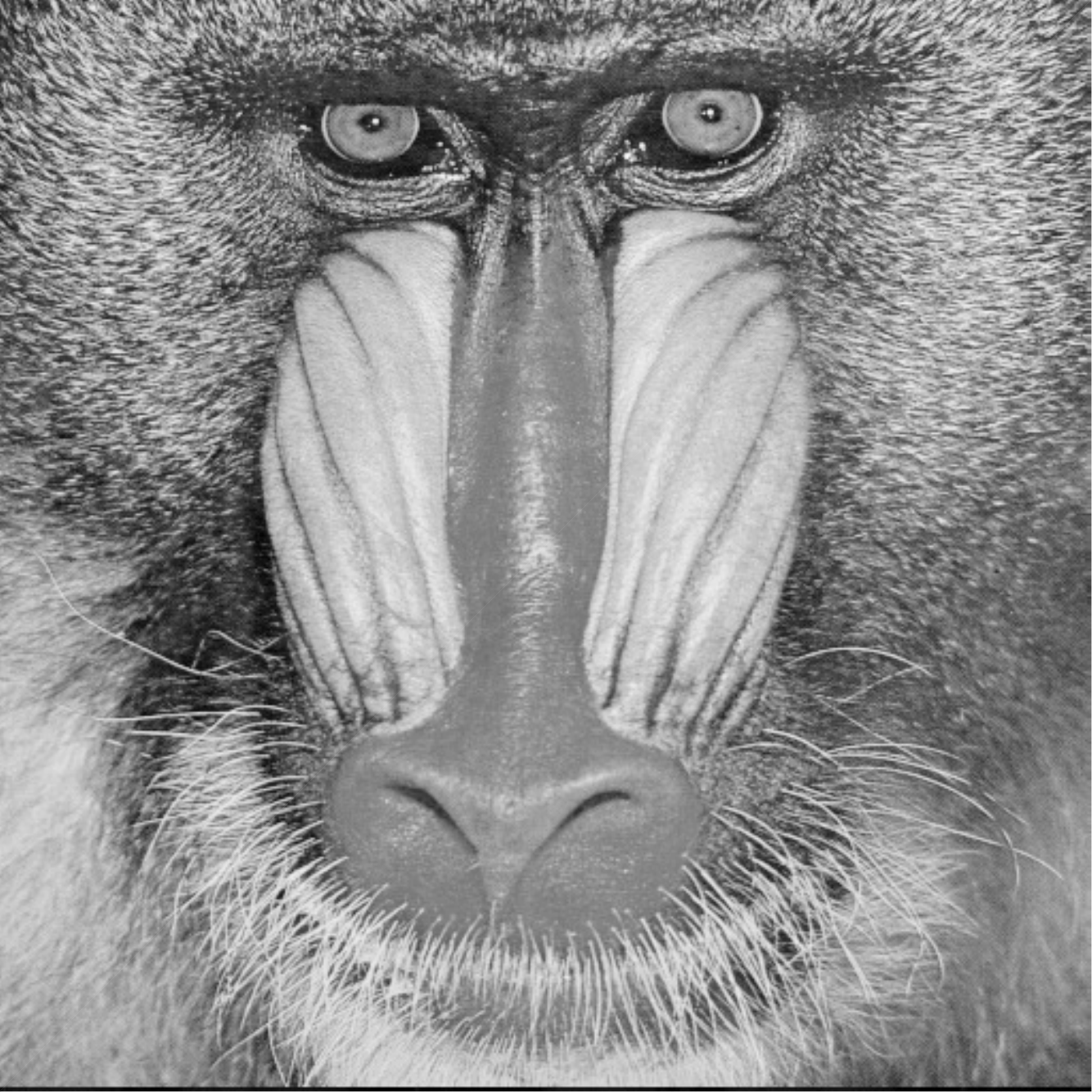}
  \caption{ }
  \label{fig7:sub1}
\end{subfigure}\hspace{.5mm} 
\begin{subfigure}{0.3\textwidth}
  \includegraphics[width=\linewidth]{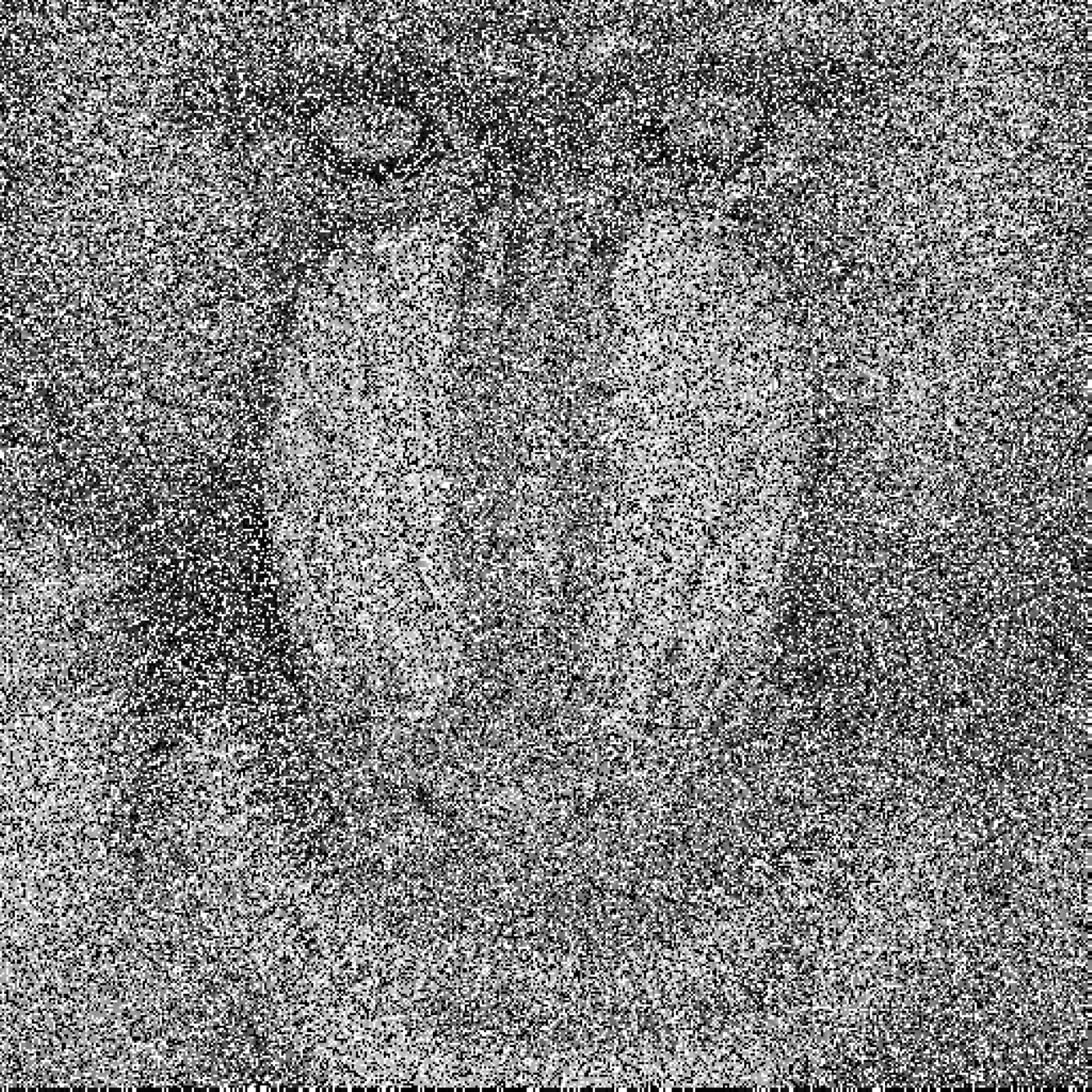}
  \caption{ }
  \label{fig7:sub2}
\end{subfigure}\hspace{.5mm} 
\begin{subfigure}{0.3\textwidth}
  \includegraphics[width=\linewidth]{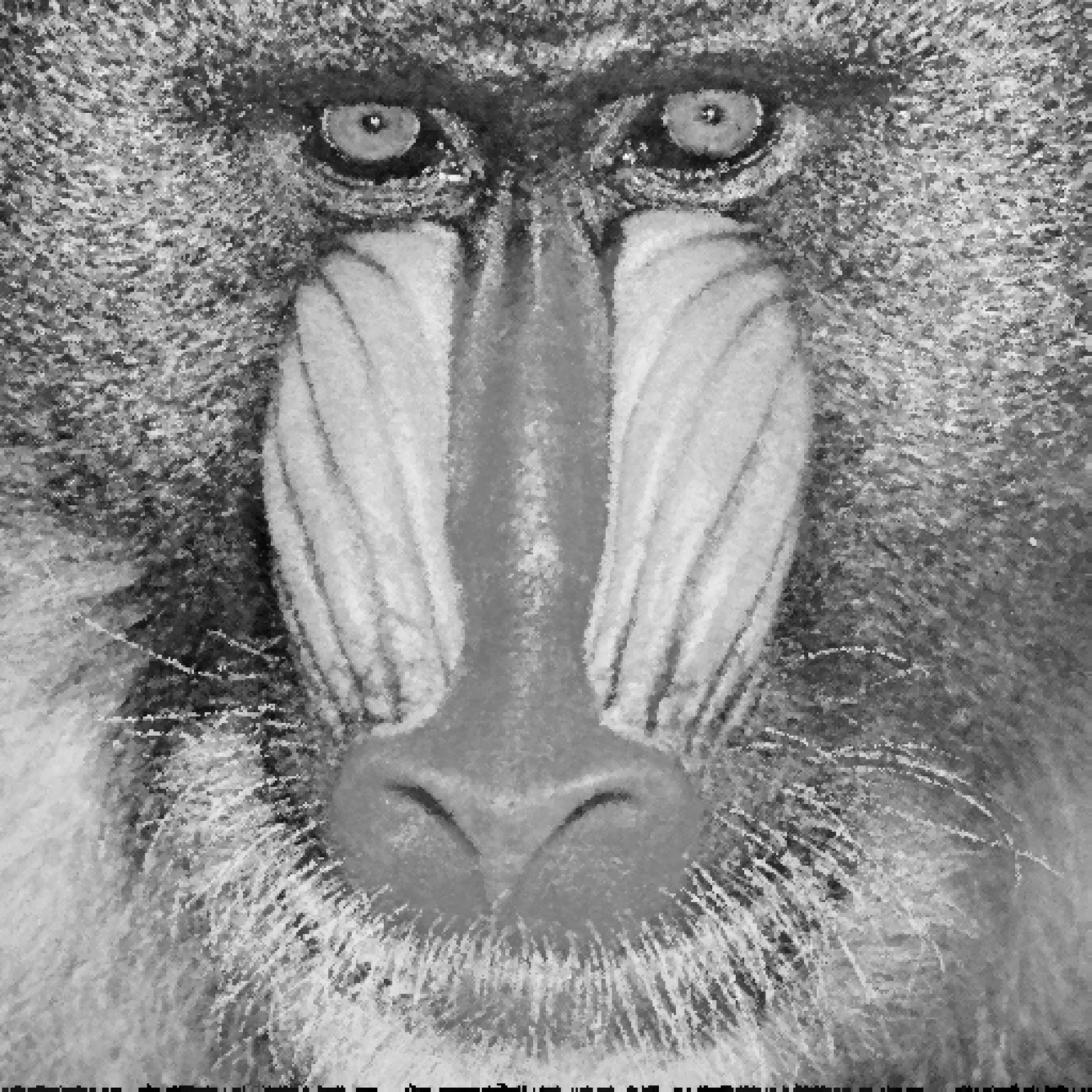}
  \caption{ }
  \label{fig7:sub3}
\end{subfigure}\hspace{.5mm} 
\begin{subfigure}{0.3\textwidth}
  \includegraphics[width=\linewidth]{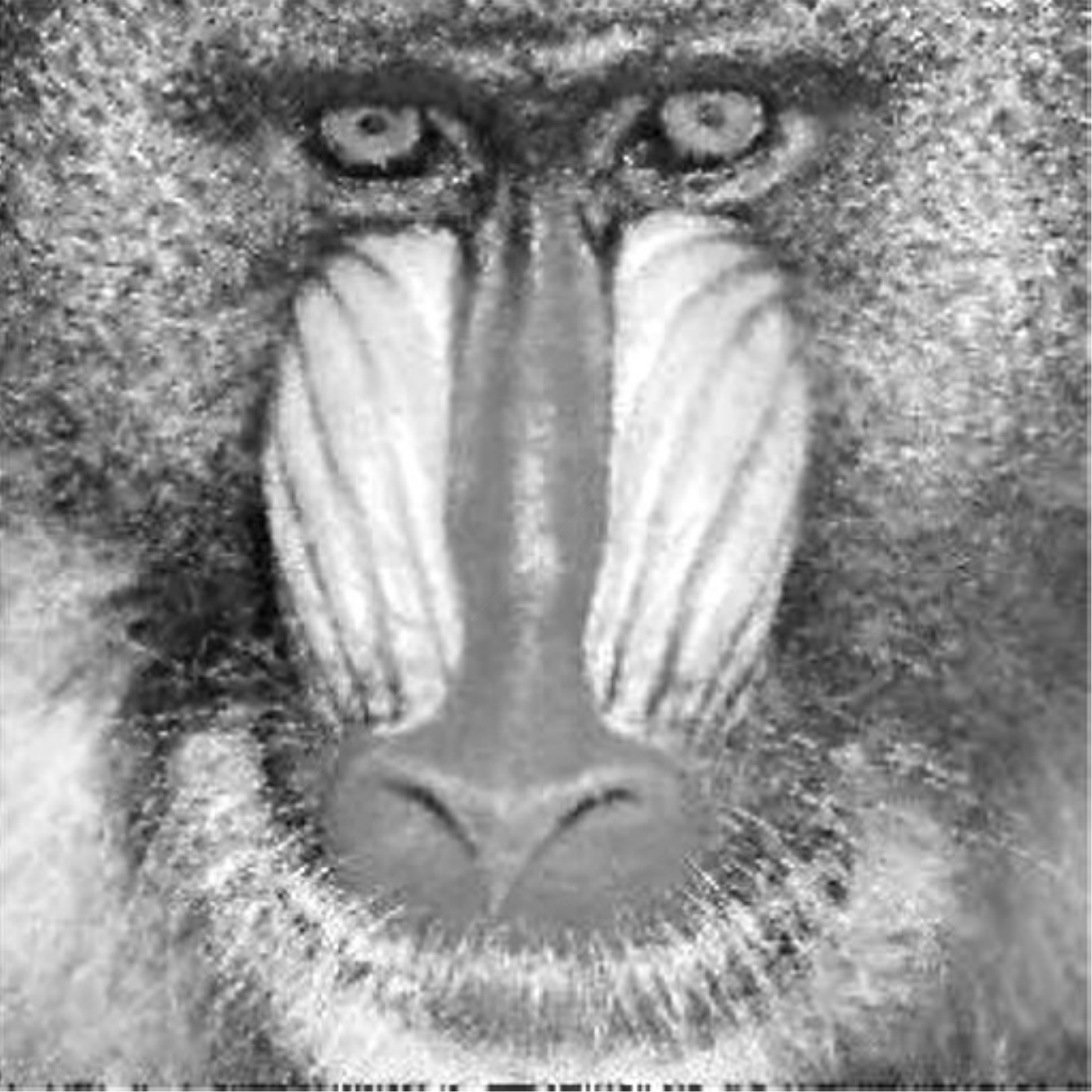}
  \caption{ }
  \label{fig7:sub4}
\end{subfigure}\hspace{.5mm} %
\begin{subfigure}{0.3\textwidth}
  \includegraphics[width=\linewidth]{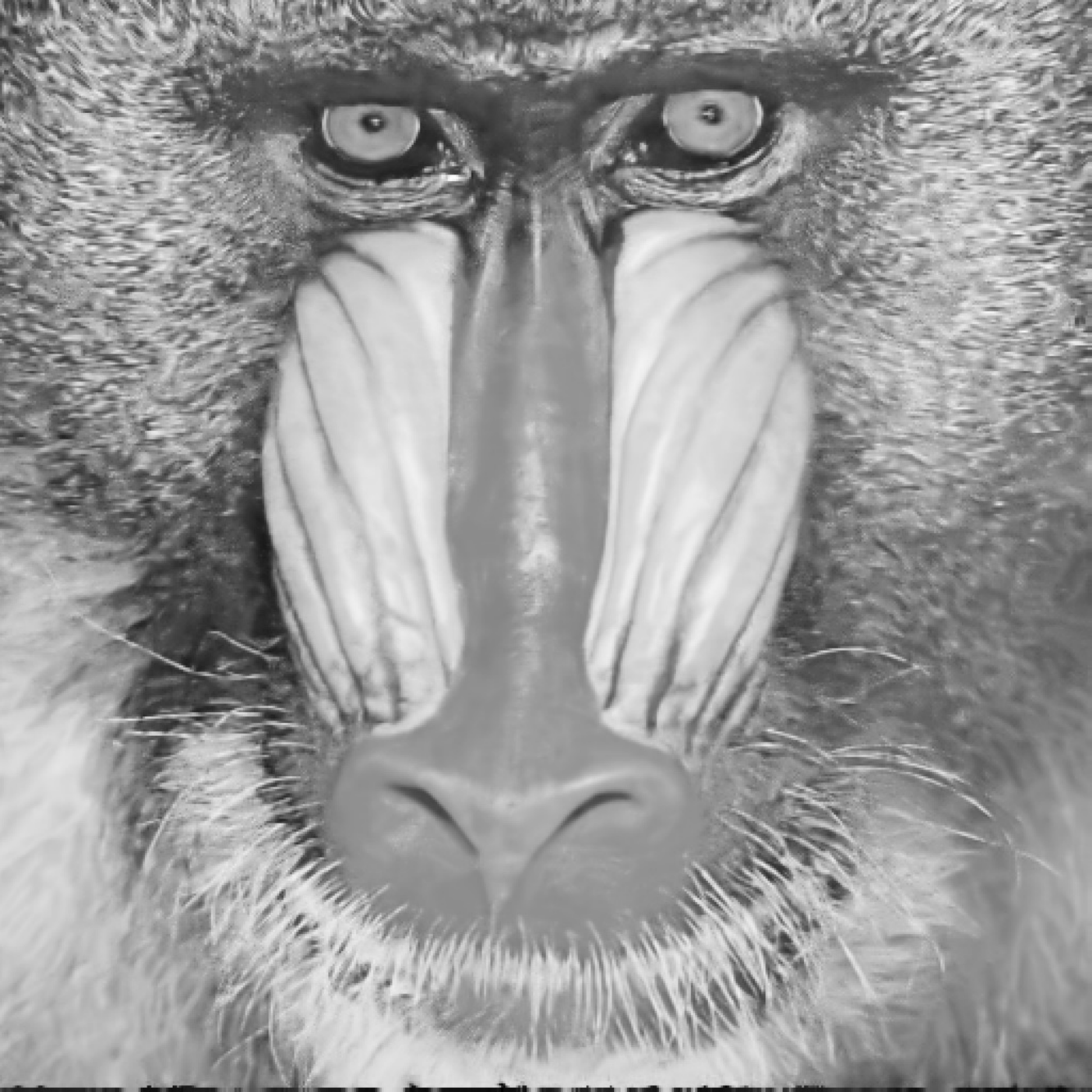}
  \caption{ }
  \label{fig7:sub5}
\end{subfigure}\hspace{.5mm} 
\begin{subfigure}{0.3\textwidth}
  \includegraphics[width=\linewidth]{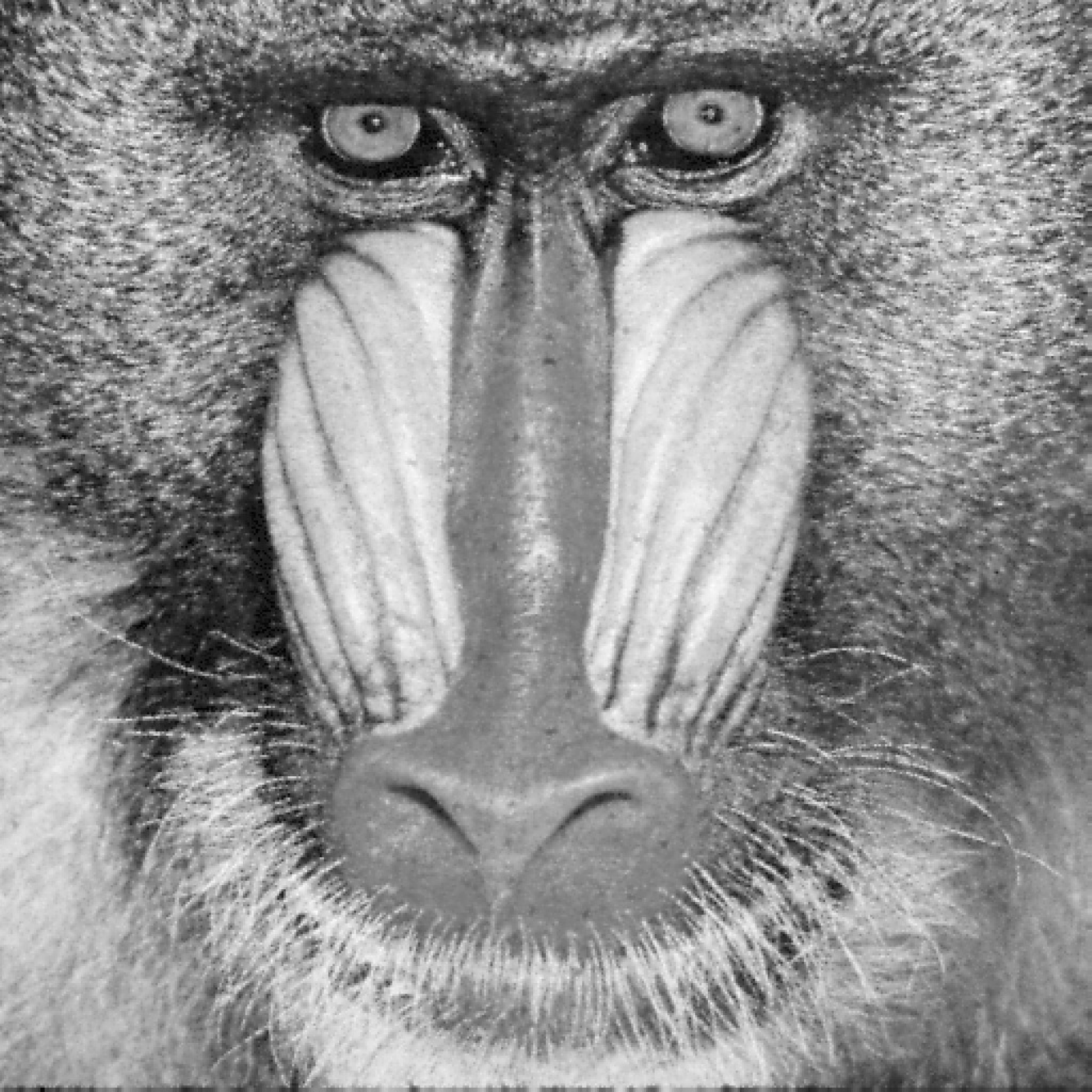}
  \caption{ }
  \label{fig7:sub6}
\end{subfigure}
\caption{Reconstructed images using different methods for the \emph{Baboon} image corrupted by 50\% SPN. (a) Original image (b) Noisy image, (c) AMF $19\times19$, (d) TPFF, (e) WESNR, (f) IDT.}
\label{fig:7}
\end{figure*}

The most common type of impulsive noise in image processing is SPN. The noisy pixels in the image corrupted by this type of noise
take the maximum or minimum values of the image, i.e., zero or 255 in 8-bit-per-pixel images. Since images are almost sparse in the DCT domain, as an example of sparse noise removal from 2-D signals, we add SPN to some images and employ Algorithm \ref{alg:2} for denoising. One should note that if we use wavelets or contourlets instead of DCT, the Peak Signal to Noise Ratio (PSNR) may be at most 0.2 dB better but the computation time is much longer. Therefore, the results of the DCT transformation are only reported. Furthermore, unknown missing samples are a special case of the SPN (the probability of salt is zero) and the quality is about the same as that of SPN.
The output PSNR for $256\times256$ cropped Lena image corrupted by 50\% SPN after each iteration of algorithm \ref{alg:2} with or without clipping plus gaussian filtering is depicted in Fig. \ref{fig:2}. As can be seen, the clipping and gaussian filtering technique result in a better reconstruction. 
We compare our results with  AMF \cite{hwang1995adaptive} with maximum window size of 19, TPFF \cite{chou2013turbulent} and WESNR \cite{jiang2014mixed}. The restoration results in terms of PSNR and Structural Similarity Metric (SSIM) \cite{sheikh2004image} for the $512\times 512$ images distorted with various densities of SPN are reported in Table \ref{tab:4}. This table shows that our algorithm outperforms the other methods in terms of SSIM and subjective evaluation for all images and noise densities. As the PSNR metric is concerned, the IDT algorithm is better than all the other methods except the WESNR, which is at most 1.5 dB better in the case of \emph{Peppers}, \emph{F-16} and \emph{Boats} images for 40\% or 50\% noise densities.
\begin{table*}
\renewcommand{\arraystretch}{1.1}
\caption{PSNR and SSIM for Different Salt-and-Pepper Noise Densities}
\label{tab:4}
\centering
\begin{tabular}{|c|c||c|c|c|c|c||c|c|c|c|c|}
    \cline{3-12}

    \multicolumn{2}{c|}{} & \multicolumn{5}{c||}{\textbf{PSNR} } & \multicolumn{5}{c|}{\textbf{SSIM} }\\
        \hline
        \cline{1-12}
        \cline{1-12}

     \multicolumn{2}{|c||}{\textbf{Noise Densities} } & \textbf{10\%} & \textbf{20\%} & \textbf{30\%} & \textbf{40\%} & \textbf{50\%}
     & \textbf{10\%} & \textbf{20\%} & \textbf{30\%} & \textbf{40\%} & \textbf{50\%} \\
        \hline
        \cline{1-12}
        \cline{1-12}

    \multirow{5}{*}{\rotatebox[origin=c]{90}{ \textbf{Lena}}} & \textbf{AMF} & 38.27 & 35.9 & 33.56 & 31.87 & 30.39
    & 0.9628 & 0.9545 & 0.9389 & 0.9174 & 0.8908  \\
    \cline{2-12}

    & \textbf{TPFF} & 35.78  & 35.06 & 32.79 & 30.98 & 29.71 & \multicolumn{5}{c|}{NOT AVAILABLE}\\
    \cline{2-12}

    & \textbf{WESNR} & 35.91  & 35.56 & 35.11 & 34.52 & 33.61  & 0.9151 & 0.9134 & 0.9099 & 0.9045 & 0.8963 \\
    \cline{2-12}

    & \textbf{IDT} & 43.35 & 39.90 & 37.56 & 35.45 & 33.54 & 0.9901 & 0.9792 & 0.9676 & 0.9527 & 0.9341 \\
    \hline
    \cline{1-12}
    \cline{1-12}

    \multirow{5}{*}{\rotatebox[origin=c]{90}{ \textbf{Peppers}}} & \textbf{AMF} & 36.06 & 33.98 & 32.17 & 30.67 & 29.23
    & 0.9482 & 0.9410 & 0.9257 & 0.9021 & 0.8733 \\
    \cline{2-12}

    & \textbf{TPFF} & 35.80  & 33.45 & 31.27 & 29.21 & 28.00 & \multicolumn{5}{c|}{NOT AVAILABLE}\\
    \cline{2-12}

    & \textbf{WESNR} & 35.01  & 34.59 & 34.08 & 33.34 & 32.49 & 0.8842 & 0.8834 & 0.8818 & 0.8785 & 0.8727 \\
    \cline{2-12}

    & \textbf{IDT} & 38.64 & 35.76 & 33.65 & 31.85 & 30.91 & 0.9811 & 0.9634 & 0.9402 & 0.9152 & 0.8891 \\
    \hline
    \cline{1-12}
    \cline{1-12}

    \multirow{5}{*}{\rotatebox[origin=c]{90}{ \textbf{F-16}}} & \textbf{AMF} & 35.87 & 32.97 & 31.05 & 29.43 & 27.77
    & 0.9780 & 0.9699 & 0.9560 & 0.9377 & 0.9118 \\
    \cline{2-12}

    & \textbf{TPFF} & 35.78  & 32.83 & 30.72 & 29.18 & 28.01 & \multicolumn{5}{c|}{NOT AVAILABLE} \\
    \cline{2-12}

    & \textbf{WESNR} & 35.27  & 34.64 & 33.96 & 32.80 & 31.85 & 0.9382 & 0.9361 & 0.9322 & 0.9264 & 0.9187 \\
    \cline{2-12}

    & \textbf{IDT} & 41.00 & 37.64 & 34.65 & 31.71 & 30.56 & 0.9814 & 0.9651 & 0.9539 & 0.9413 & 0.9284 \\
    \hline
    \cline{1-12}
    \cline{1-12}

     \multirow{5}{*}{\rotatebox[origin=c]{90}{ \textbf{Baboon}}} & \textbf{AMF} & 26.95 & 25.73 & 24.53 & 23.29 & 22.14
     & 0.8922 & 0.8717 & 0.8351 & 0.7866 & 0.7264 \\
    \cline{2-12}

     & \textbf{TPFF} & 30.96  & 27.90 & 26.34 & 25.15 & 23.87 & \multicolumn{5}{c|}{NOT AVAILABLE}  \\
    \cline{2-12}

    & \textbf{WESNR} & 26.44  & 26.17 & 25.70 & 24.93 & 24.11 & 0.7982 & 0.7938 & 0.7784 & 0.7529 & 0.7170 \\
    \cline{2-12}

    & \textbf{IDT} & 32.41 & 29.24 & 27.17 & 25.60 & 24.38 & 0.9751 & 0.9449 & 0.9088 & 0.8654 & 0.8116 \\
    \hline
    \cline{1-12}
    \cline{1-12}

     \multirow{4}{*}{\rotatebox[origin=c]{90}{ \textbf{Boat}}} & \textbf{AMF} & 33.94 & 32.05 & 30.25 & 28.74 & 27.16
     & 0.9345 & 0.9215 & 0.8986 & 0.8679 & 0.8274 \\
    \cline{2-12}

    & \textbf{TPFF} & \multicolumn{10}{c|}{NOT AVAILABLE}\\
    \cline{2-12}

    & \textbf{WESNR} & 32.78 & 32.40 & 31.84 & 31.13 & 30.08 & 0.8659 & 0.8635 & 0.8567 & 0.8478 & 0.8317 \\
    \cline{2-12}

    & \textbf{IDT} & 37.91 & 34.91 & 32.68 & 30.77 & 29.15 & 0.9791 & 0.9579 & 0.9340 & 0.9082 & 0.8744 \\
    \hline

\end{tabular}
\end{table*}
Figure \ref{fig:7} exhibits the restored images of various methods for the \emph{Baboon} image corrupted by 50\% SPN. This figure shows that the TPFF and AMF algorithms are not capable of completely removing the noise and the WESNR algorithm smoothes the image but our proposed method preserves the details of the image and remove the noise completely.
%
%

\begin{figure*}
    \centering 
\begin{subfigure}{0.3\textwidth}
  \includegraphics[width=\linewidth]{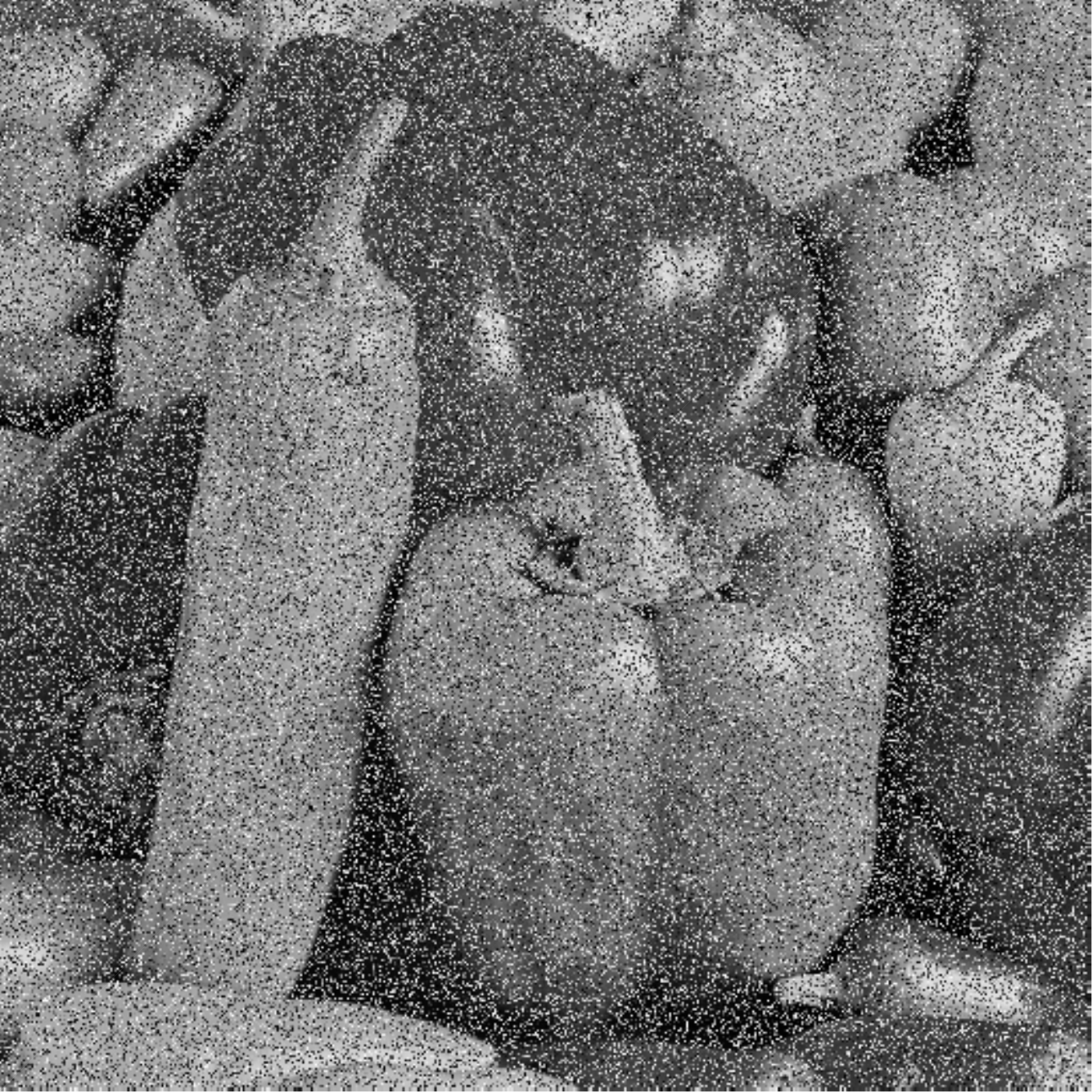}
  \caption{ }
  \label{fig8:sub1}
\end{subfigure}\hspace{.5mm} 
\begin{subfigure}{0.3\textwidth}
  \includegraphics[width=\linewidth]{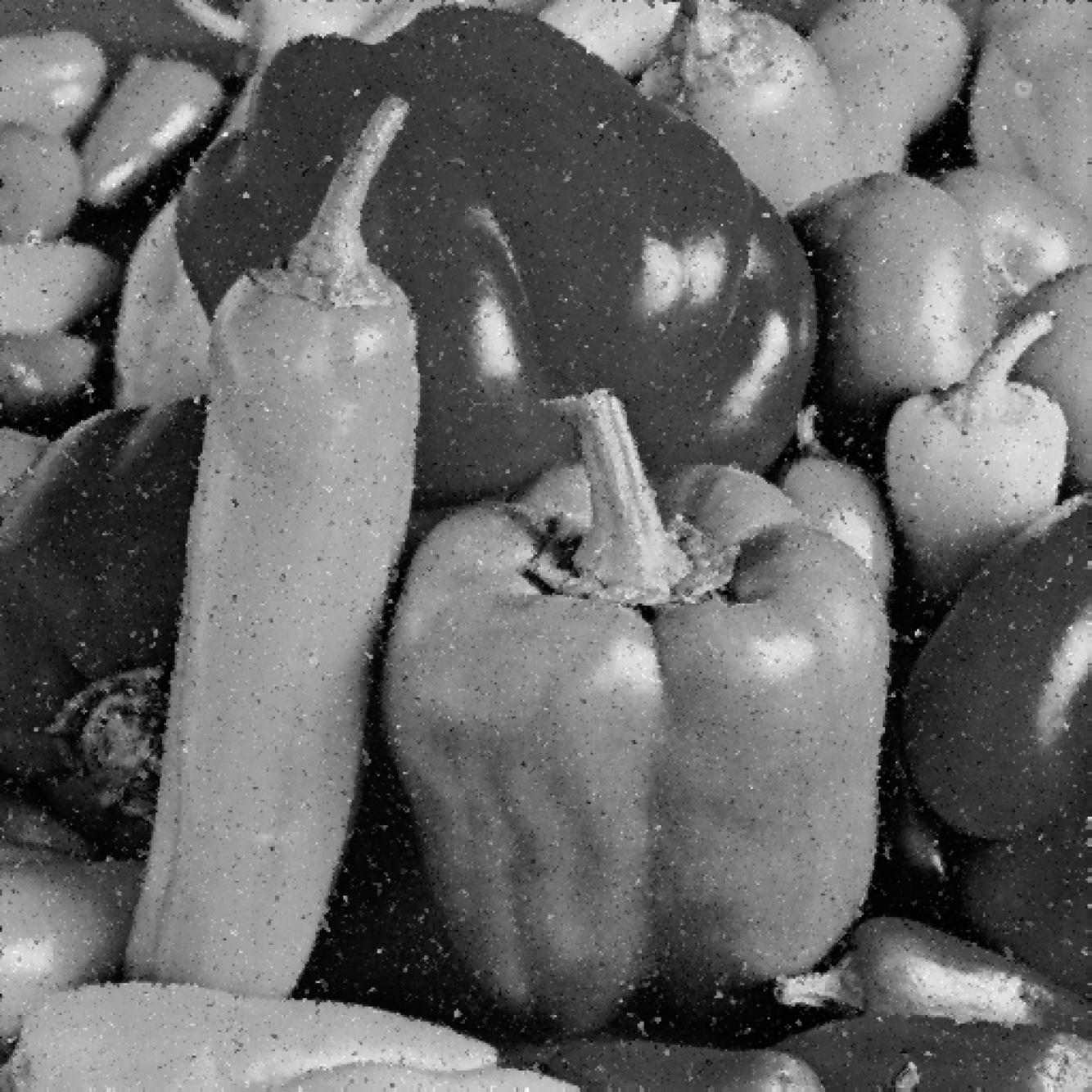}
  \caption{ }
  \label{fig8:sub2}
\end{subfigure}\hspace{.5mm} %
\begin{subfigure}{0.3\textwidth}
  \includegraphics[width=\linewidth]{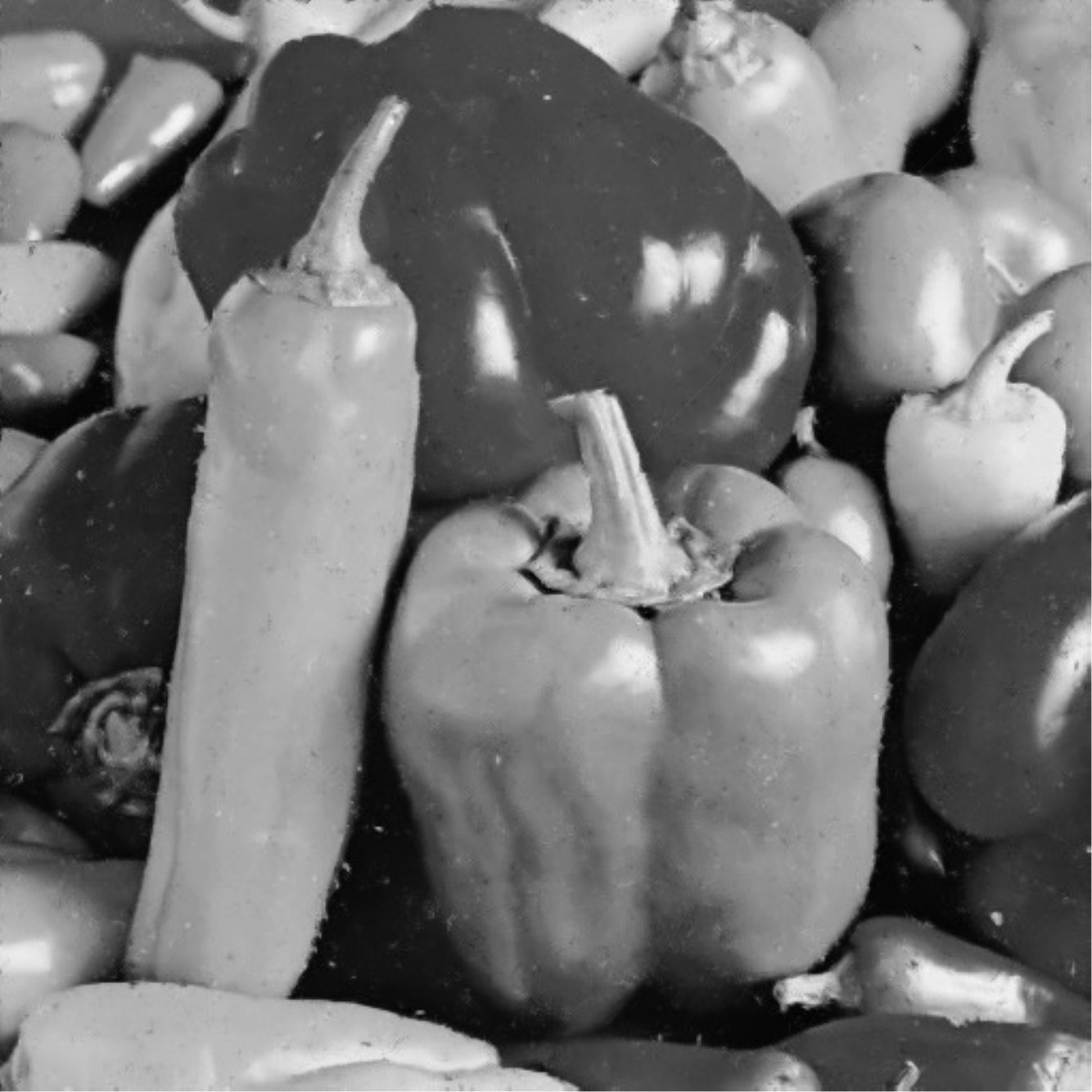}
  \caption{ }
  \label{fig8:sub3}
\end{subfigure}\hspace{.5mm} 
\begin{subfigure}{0.3\textwidth}
  \includegraphics[width=\linewidth]{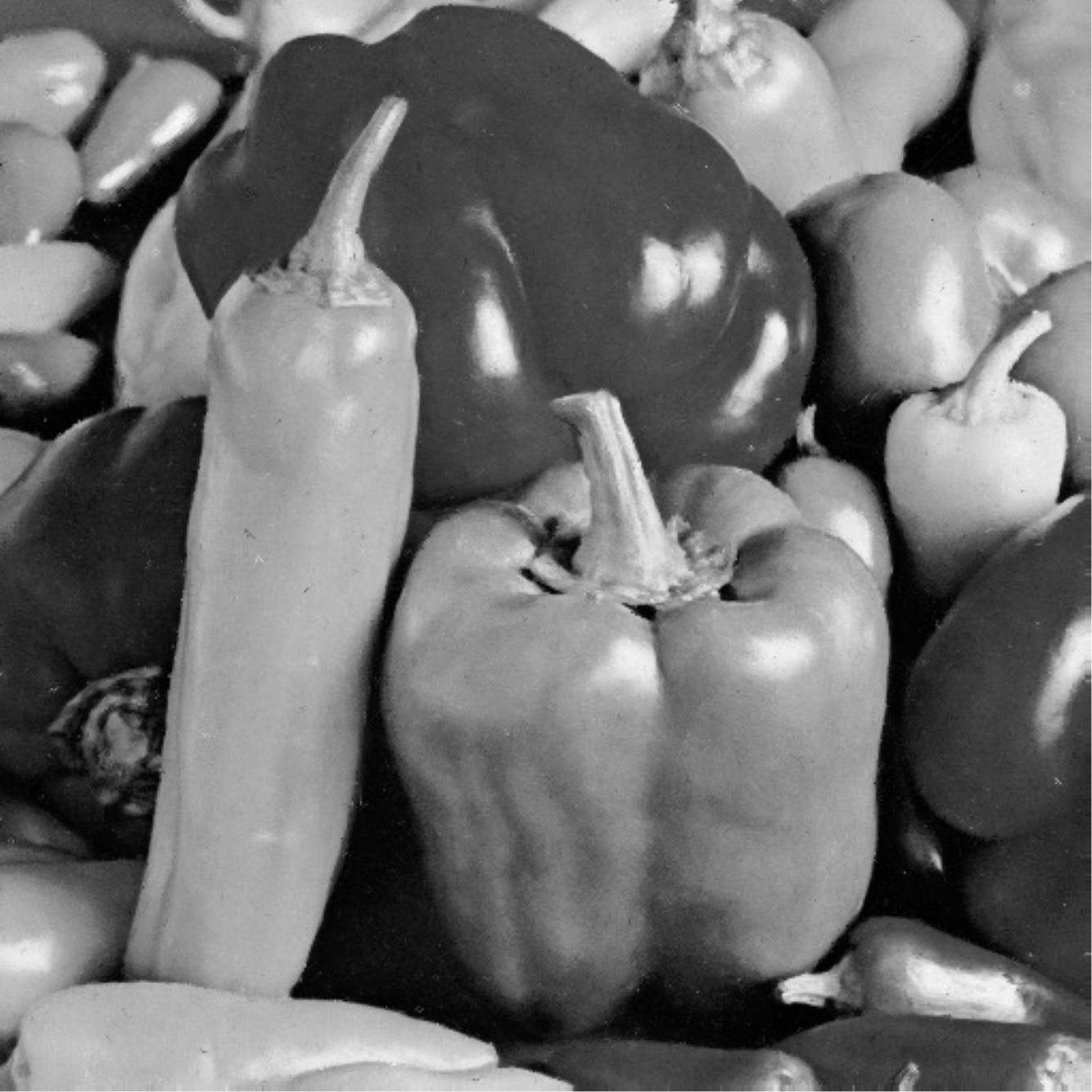}
  \caption{ }
  \label{fig8:sub4}
\end{subfigure}\hspace{.5mm} 
\begin{subfigure}{0.3\textwidth}
  \includegraphics[width=\linewidth]{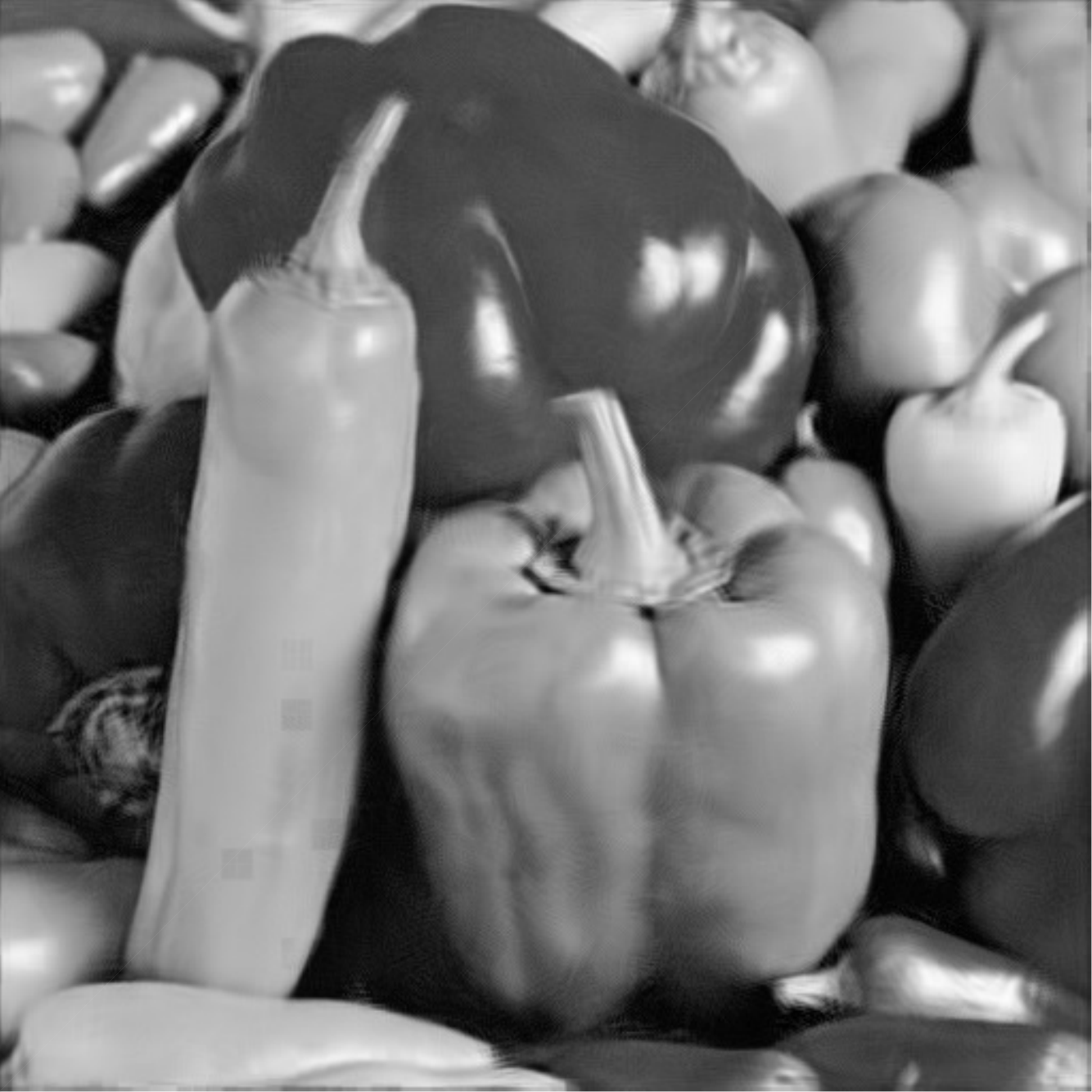}
  \caption{ }
  \label{fig8:sub5}
\end{subfigure}\hspace{.5mm} %
\begin{subfigure}{0.3\textwidth}
  \includegraphics[width=\linewidth]{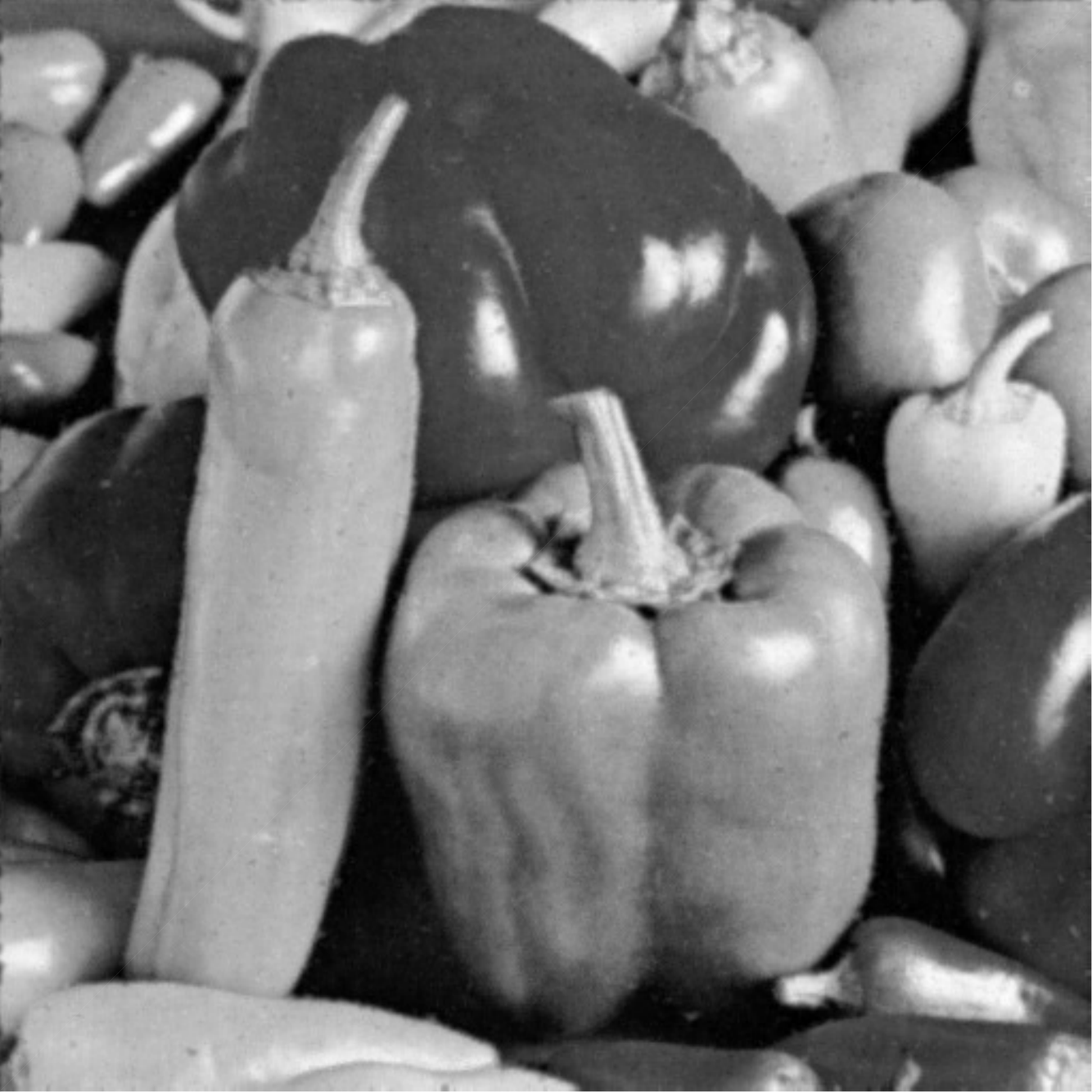}
  \caption{ }
  \label{fig8:sub6}
\end{subfigure}
\caption{Reconstructed images using different methods for the \emph{Peppers} image corrupted by 40\% random-valued impulsive noise.(a) Noisy image, (b) ACWMF, (c) WESNR, (d) SAFE, (e) ALOHA, (f) IDT.}
\label{fig:8}
\end{figure*}

\begin{figure}
\centering%
\begin{subfigure}{0.48\linewidth}
  \centering
  \includegraphics[width=0.95\linewidth]{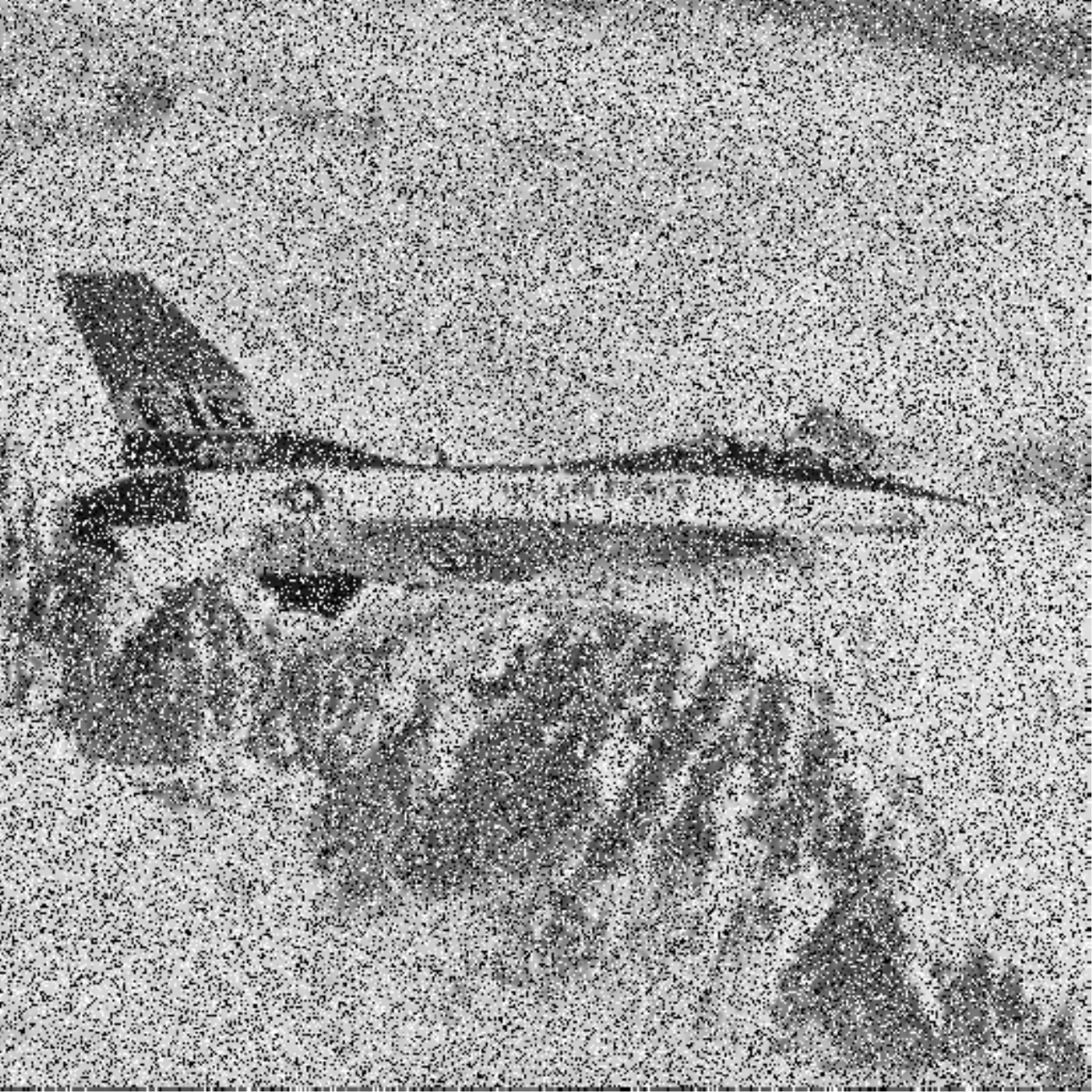}
  \caption{10.06 dB}
  \label{fig6:sub1}
\end{subfigure}\hspace{.3mm}%
\begin{subfigure}{0.48\linewidth}
  \centering
  \includegraphics[width=0.95\linewidth]{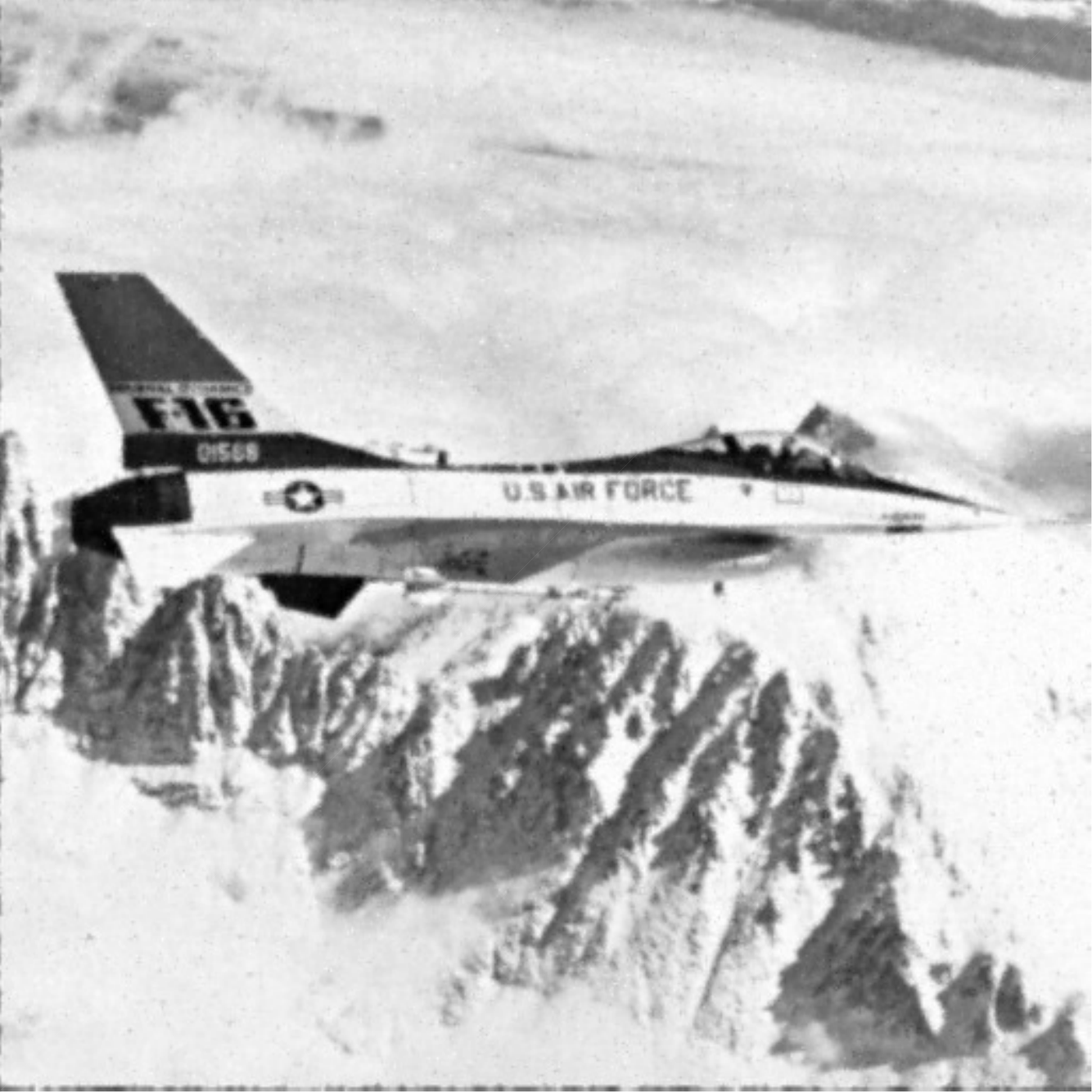}
  \caption{33.31 dB}
  \label{fig6:sub2}
\end{subfigure}
\caption{Restored images for the \emph{F-16} image corrupted by 15\% RVIN and 25\% SPN. (a) Noisy image, (b) Restored image.}\label{fig:6}
\end{figure}

\begin{figure}
\centering%
\begin{subfigure}{0.48\linewidth}
  \centering
  \includegraphics[width=0.95\linewidth]{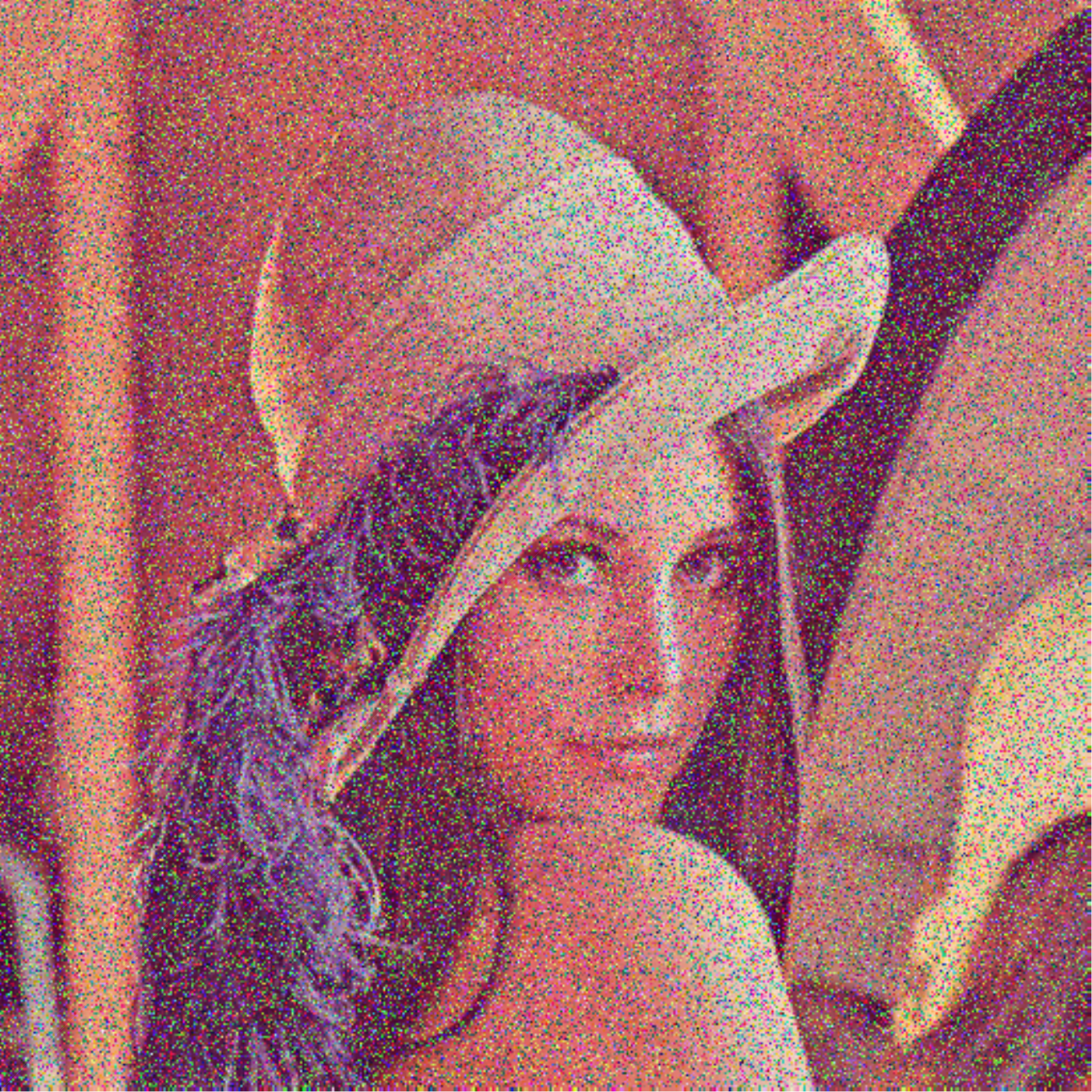}
  \caption{}
  \label{fig5:sub1}
\end{subfigure}\hspace{.3mm}%
\begin{subfigure}{0.48\linewidth}
  \centering
  \includegraphics[width=0.95\linewidth]{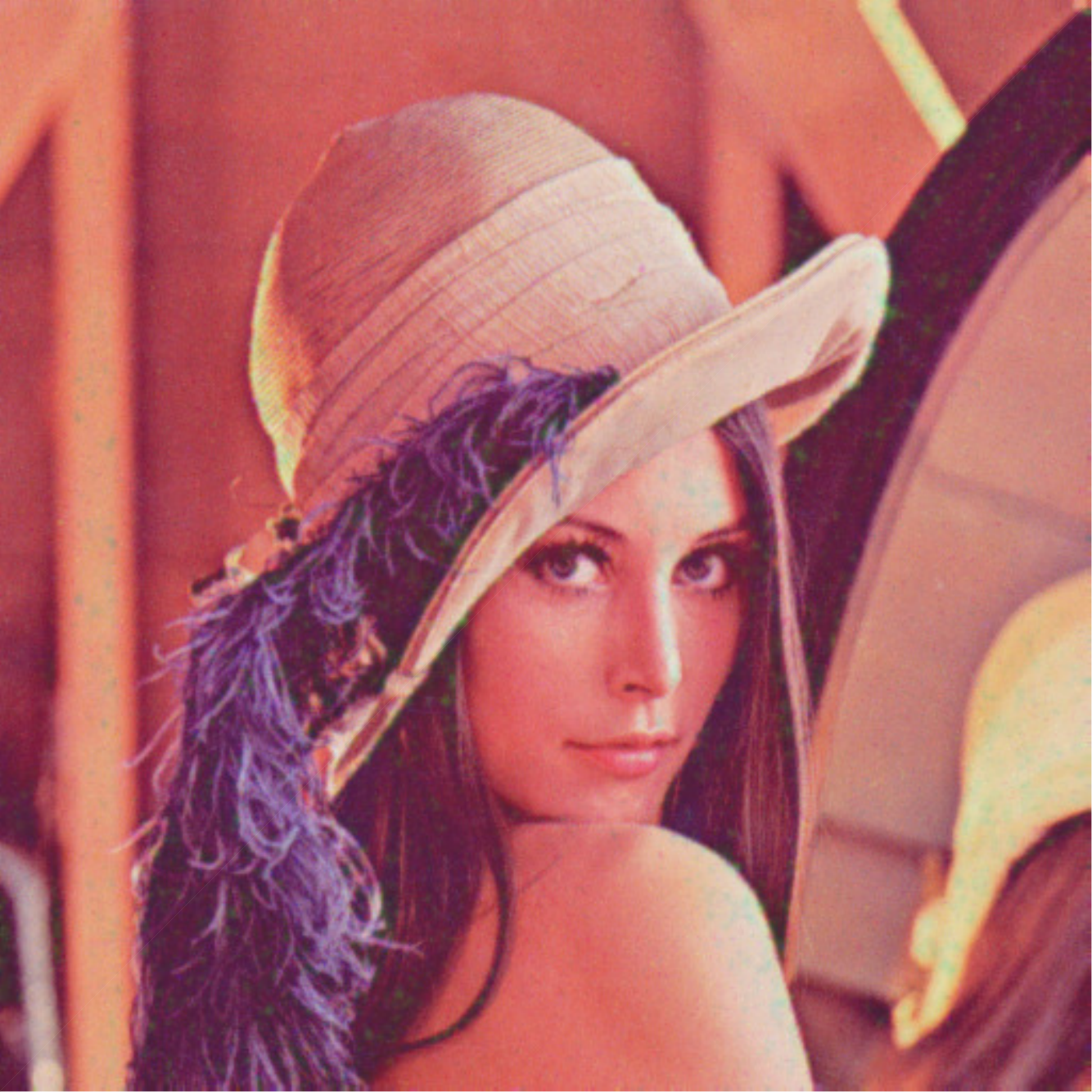}
  \caption{}
  \label{fig5:sub2}
\end{subfigure}
\caption{Restored images for the colored \emph{Lena} image corrupted by 30\% RVIN. (a) Noisy image, (b) Restored image.}\label{fig:5}
\end{figure}

As another example of sparse noise, we consider RVIN. In this case, each noisy pixel randomly takes a value in the interval $[0,255]$ with uniform distribution. We compare our algorithm with the methods called ACWMF \cite{chen2001adaptive}, WESNR \cite{jiang2014mixed}, SAFE \cite{chen2016structure} and ALOHA \cite{jin2018sparse}. The results can be found in Table \ref{tab:3}.
\begin{table*}
\renewcommand{\arraystretch}{1.1}
\caption{PSNR and SSIM for Different Random-Valued Impulsive Noise Densities (ND)}
\label{tab:3}
\centering
\begin{tabular}{|c|c||c|c|c|c|c|c||c|c|c|c|c|c|}
    \cline{3-14}

    \multicolumn{2}{c|}{} & \multicolumn{6}{c||}{\textbf{PSNR} } & \multicolumn{6}{c|}{\textbf{SSIM} }\\
        \hline
        \cline{1-14}
        \cline{1-14}

     \multicolumn{2}{|c||}{\textbf{Noise Densities} } & \textbf{5\%} & \textbf{10\%} & \textbf{20\%} & \textbf{30\%} & \textbf{40\%} & \textbf{50\%}
     & \textbf{5\%} &  \textbf{10\%} & \textbf{20\%} & \textbf{30\%} & \textbf{40\%} & \textbf{50\%} \\
        \hline
        \cline{1-14}
        \cline{1-14}

    \multirow{6}{*}{\rotatebox[origin=c]{90}{ \textbf{Lena}}} & \textbf{ACWMF} & 38.53 & 35.32 & 31.61 & 28.76 & 26.15 & 23.52
    & 0.9669 & 0.9325 & 0.8562 & 0.7592 & 0.6459 & 0.5133   \\
    \cline{2-14}

    & \textbf{WESNR} & 36.83 & 36.30 & 35.10 & 33.00 & 30.91 & 28.18 & 0.9271 & 0.9245 & 0.9165 & 0.8972 & 0.8619 & 0.7880 \\
    \cline{2-14}

    & \textbf{SAFE} & 34.97 & 35.92 & 34.79 & 33.33 & 31.97 & 30.43 & 0.9587 & 0.9635 & 0.9551 & 0.9407 & 0.9208 & 0.8930 \\
    \cline{2-14}

    & \textbf{ALOHA} & 40.79 & 38.81 & 35.32 & 32.66 & 30.62 & 26.69 &  0.9821 &  0.9695 &  0.9403 &  0.9023 &  0.8752 &  0.7729 \\
    \cline{2-14}

    & \textbf{IDT} & 40.10 & 37.75 & 34.88 & 32.97 & 31.34 & 29.73 & 0.9839 & 0.9732 & 0.9516 & 0.9293 & 0.9023 & 0.8712 \\
    \hline
    \cline{1-14}
    \cline{1-14}

    \multirow{6}{*}{\rotatebox[origin=c]{90}{ \textbf{Peppers}}} & \textbf{ACWMF} & 36.73 & 34.30 & 30.56 & 27.85 & 25.06 & 22.23
    & 0.9659 & 0.9311 & 0.8494 & 0.7469 & 0.6201 & 0.4819 \\
    \cline{2-14}

    & \textbf{WESNR} & 35.31 & 34.68 & 33.61 & 31.83 & 29.13 & 26.01 & 0.8955 & 0.8949 & 0.8888 & 0.8752 & 0.8342 & 0.7398 \\
    \cline{2-14}

    & \textbf{SAFE} & 30.56 & 30.80 & 30.40 & 29.82 & 28.97 & 28.40 & 0.9492 & 0.9537 & 0.9423 & 0.9151 & 0.8742 & 0.8812 \\
    \cline{2-14}

    & \textbf{ALOHA} & 37.55 &  35.91 &  33.00 &  31.16 &  28.50 &  25.02 &  0.9585 & 0.9310 & 0.8759 & 0.8337 & 0.7767 & 0.6824 \\
    \cline{2-14}

    & \textbf{IDT} & 37.40 & 35.20 & 32.10 & 31.00 & 29.53 & 27.92 & 0.9770 & 0.9653 & 0.9303 & 0.9151 & 0.8850 & 0.8461 \\
    \hline
    \cline{1-14}
    \cline{1-14}

    \multirow{6}{*}{\rotatebox[origin=c]{90}{ \textbf{F-16}}} & \textbf{ACWMF} & 36.78 & 33.62 & 29.95 & 27.28 & 24.57 & 21.72
    & 0.9700 & 0.9351 & 0.8561 & 0.7576 & 0.6220 & 0.4784 \\
    \cline{2-14}

    & \textbf{WESNR} & 35.75 & 34.80 & 32.85 & 30.72 & 28.37 & 25.44 & 0.9453 & 0.9420 & 0.9321 & 0.9115 & 0.8601 & 0.7511 \\
    \cline{2-14}

    & \textbf{SAFE} & 28.29 & 28.81 & 28.64 & 28.02 & 27.31 & 26.47 & 0.9498 & 0.9525 & 0.9447 & 0.9188 & 0.8804 & 0.8898 \\
    \cline{2-14}

    & \textbf{ALOHA} & 36.17 & 34.93 & 32.26 & 28.33 & 27.76 & 24.68 &  0.9410 & 0.9255 & 0.9072 & 0.8788 & 0.8633 & 0.7503 \\
    \cline{2-14}

    & \textbf{IDT} & 38.28 & 35.77 & 32.75 & 30.65 & 28.81 & 27.05 & 0.9853 & 0.9771 & 0.9595 & 0.9382 & 0.9113 & 0.8741 \\
    \hline
    \cline{1-14}
    \cline{1-14}

     \multirow{6}{*}{\rotatebox[origin=c]{90}{ \textbf{Baboon}}} & \textbf{ACWMF} & 27.63 & 26.31 & 24.37 & 22.64 & 21.28 & 19.83
     & 0.9311 & 0.8955 & 0.8214 & 0.7299 & 0.6329 & 0.5171 \\
    \cline{2-14}

     & \textbf{WESNR} & 26.90 & 26.12 & 24.86 & 23.76 & 22.62 & 21.42 & 0.8420 & 0.8251 & 0.7852 & 0.7402 & 0.6761 & 0.5931  \\
    \cline{2-14}

    & \textbf{SAFE} & 24.02 & 24.57 & 24.26 & 23.35 & 22.24 & 21.19 & 0.8217 & 0.8558 & 0.8382 & 0.7827 & 0.7126 & 0.6305 \\
    \cline{2-14}

    & \textbf{ALOHA} & 29.37 &  28.59 &  25.56 &  22.66 &  22.33 &  20.56 &  0.7721 & 0.7716 & 0.7015 & 0.6495 & 0.6433 & 0.5792 \\
    \cline{2-14}

    & \textbf{IDT} & 30.86 & 28.26 & 25.50 & 23.83 & 22.50 & 21.43 & 0.9446 & 0.9061 & 0.8340 & 0.7604 & 0.6800 & 0.6069 \\
    \hline
    \cline{1-14}
    \cline{1-14}

     \multirow{6}{*}{\rotatebox[origin=c]{90}{ \textbf{Boat}}} & \textbf{ACWMF} & 34.65 & 32.29 & 29.15 & 26.86 & 24.64 & 22.31
     & 0.9633 & 0.9274 & 0.8476 & 0.7545 & 0.6451 & 0.5193 \\
    \cline{2-14}

    & \textbf{WESNR} & 33.13 & 32.55 & 31.24 & 29.58 & 27.85 & 25.72 & 0.8892 & 0.8835 & 0.8675 & 0.8387 & 0.7945 & 0.7161 \\
    \cline{2-14}

    & \textbf{SAFE} & 30.19 & 31.33 & 30.65 & 29.26 & 28.04 & 26.62 & 0.9171 & 0.9377 & 0.9187 & 0.8922 & 0.8560 & 0.8051 \\
    \cline{2-14}

    & \textbf{ALOHA} & 35.88 & 34.53 & 31.11 & 29.47 & 26.99 & 24.29 & 0.9574 & 0.9536 & 0.8869 & 0.8571 & 0.7594 & 0.6818 \\
    \cline{2-14}

    & \textbf{IDT} & 35.02 & 33.01 & 30.58 & 28.91 & 27.65 & 25.88 & 0.9688 & 0.9505 & 0.9148 & 0.8781 & 0.8371 & 0.7876 \\
    \hline

\end{tabular}
\end{table*}
Our algorithm is better than or equal to the other methods in terms of the SSIM when the noise density is less than 30\%, but the SAFE algorithm is slightly better for higher noise densities. Moreover, the IDT algorithm outperforms all the other methods for the \emph{F-16} and \emph{Baboon} images and all RVIN densities in terms of PSNR. For other images, from PSNR point of view, our algorithm is more or less the same as the ALOHA algorithm when the noise density is less than 30\%, and for higher noise densities, the IDT algorithm is more or less the same as the SAFE algorithm. The ALOHA and the SAFE algorithms do not work well for SPN removal, and it is demonstrated in the next subsection that the run-time of our algorithm is much less than these two methods. The restored images for the \emph{peppers} image corrupted by 40\% RVIN are depicted in Figure \ref{fig:8}. This figure shows that the ACWMF and the WESNR algorithms are not capable of removing all the impulsive noise. The ALOHA algorithm adds some artifacts to the image and the SAFE and the proposed algorithm have the same subjective quality.
Our algorithm also performs well when the image is corrupted by mixed RVIN and SPN. The results for the \emph{F-16} image corrupted by 15\% SPN and 25\% RVIN can be found in Figure \ref{fig:6}.
The proposed algorithm can easily be applied to colored images by denoising each channel separately. As an example, the colored \emph{Lena} image is corrupted by 30\% RVIN and the reconstructed image is depicted in \ref{fig:5}.

\subsection{Audio Reconstruction}
\begin{figure*}
    \centering 
\begin{subfigure}{0.48\textwidth}
  \includegraphics[width=\linewidth]{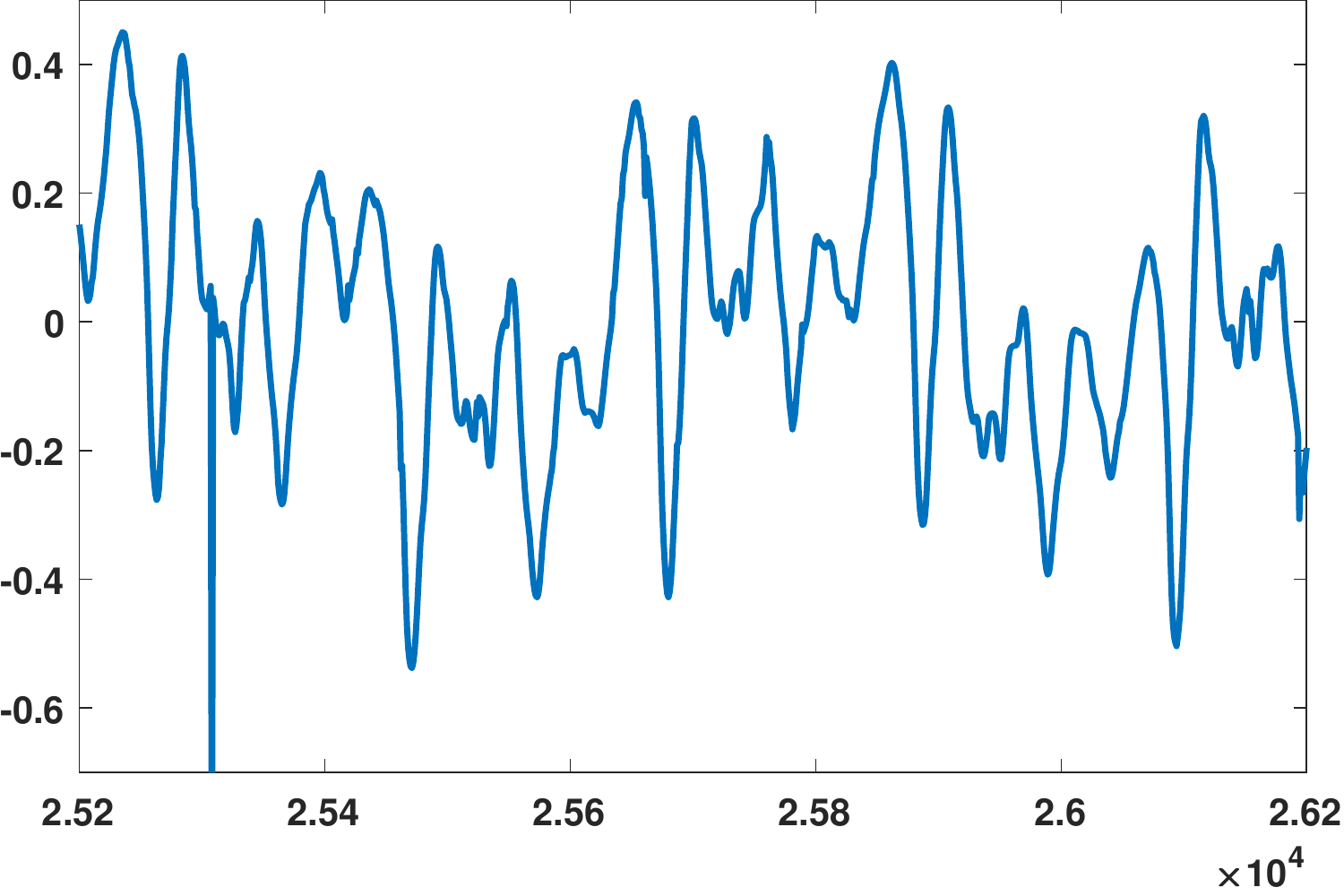}
  \caption{ }
  \label{fig1:sub2}
\end{subfigure}\hspace{.1mm} 
\begin{subfigure}{0.48\textwidth}
  \includegraphics[width=\linewidth]{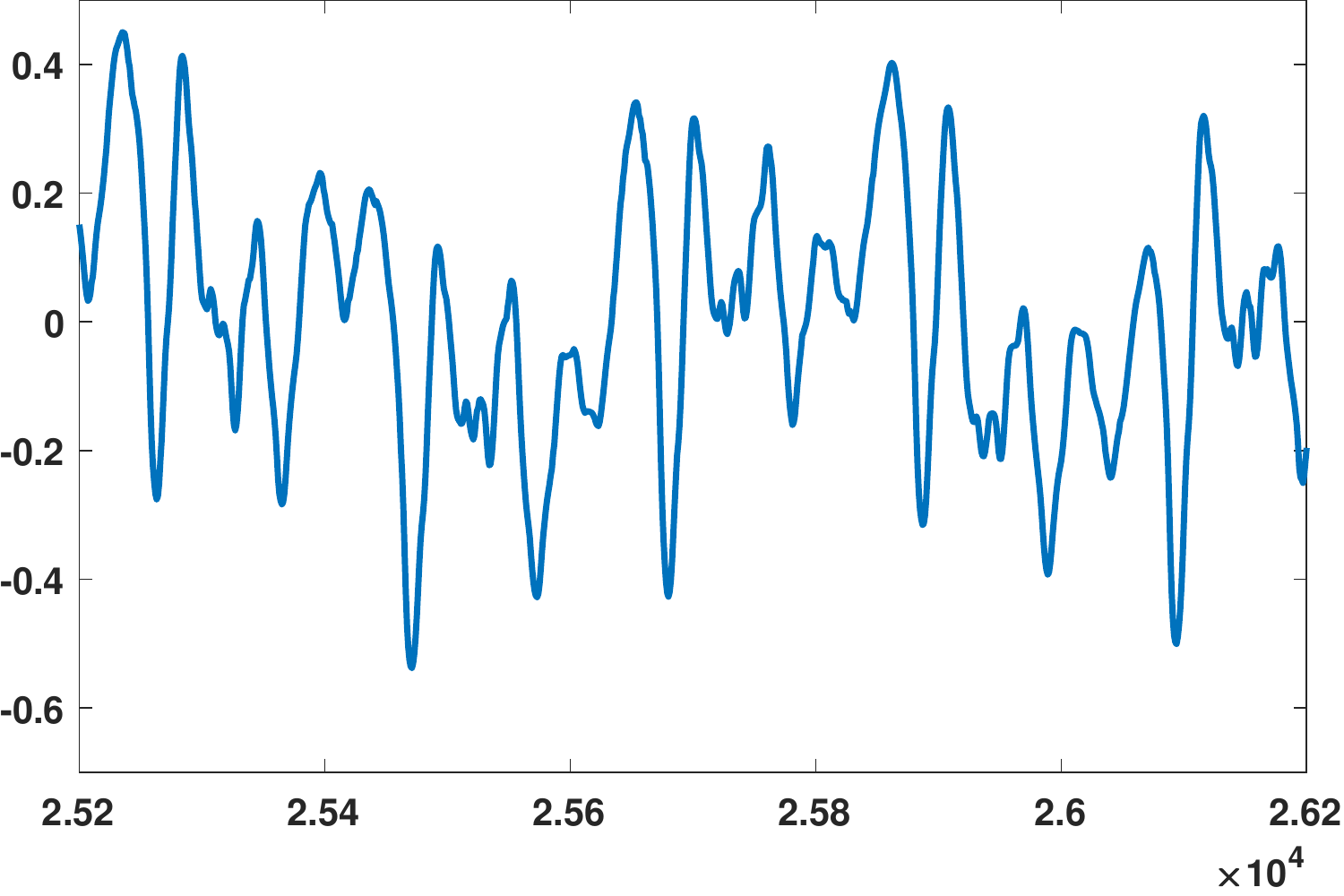}
  \caption{ }
  \label{fig1:sub3}
\end{subfigure}\hspace{.1mm}
\caption{Recovery of audio degraded with clicks. (a) Degraded audio (SNR = 26.27 dB, PEAQ = 2.45), (b) Restored audio (SNR = 35.55, PEAQ = 1.11).}
\label{fig:1}
\end{figure*}

Clicks and pops are localized bursts of impulsive noise in audio signals and they are commonly caused by particles or scratches on the surface of a phonograph record or CD. In the recent publications in impulsive noise removal from audio signals \cite{stankovic2018analysis,ciolek2017detection}, it is assumed that the noisy samples are detected beforehand, and thus it is not fair to compare our results with theirs. As an example of impulsive noise removal from 1-D signals, we use IDT algorithm to remove clicks from a 5 second of the country music \emph{'you are my sunshine'} \cite{nuzman2004audio}. The noisy and reconstructed audio signals are presented in Figure \ref{fig:1}. The SNR and Perceptual Evaluation of Audio Quality (PEAQ) are given in the same figure. For the quality of the restored audio signal, please check 'acri.ee.sharif.ir'.

\subsection{Complexity}\label{complexity}
We evaluate the complexity of our method in terms of the run-time in this subsection. The simulation is conducted for the \emph{Lena} image corrupted with 30\% densities of impulsive noise. The results are shown in Table \ref{tab:2}.
\begin{table}
\renewcommand{\arraystretch}{1.1}
\caption{Run-Time of Different Methods in Seconds}
\label{tab:2}
\centering
\setlength\tabcolsep{5pt}
\begin{tabular}{|c|c|c|c|c|c|c|}
    \hline

    \multirow{2}{*}{\textbf{AMF}}  & \multirow{2}{*}{\textbf{ACWMF}} & \multirow{2}{*}{\textbf{TPFF}} & \multirow{2}{*}{\textbf{WESNR}} & \textbf{SAFE} & \textbf{ALOHA} & \multirow{2}{*}{\textbf{IDT}}\\
      &  &  &  & \textbf{(on GPU)} & \textbf{on GPU} & \\

    \cline{1-7}
    \cline{1-7}

    \multirow{2}{*}{1.13} & \multirow{2}{*}{1.94} & \multirow{2}{*}{1.75} & \multirow{2}{*}{34.32} & 763.48 & \multirow{2}{*}{1368.2} & \multirow{2}{*}{1.71}\\
      &  &  &  & (21.07) & & \\
    \hline

\end{tabular}
\end{table}
%
%
%
%
%
%
%
%
This table shows that our method is very efficient and fast. On the other hand, the SAFE and ALOHA algorithms are extremely inefficient while the WESNR method is about 20 times slower than IDT. In each iteration of the proposed method, there are one 2D-DCT, one 2D-IDCT, two thresholding, one clipping and one gaussian filtering. The dominant terms in computational complexity are 2D-DCT and 2D-IDCT, and hence the overall complexity of our algorithm is $\mathcal{O}(n^{2})\log n$ \cite{ahmed1974discrete}.

\section{Conclusion}\label{conclusion}
In this paper, we proposed a new method for separating two signal that are sparse in two different domains. Among the applications of this problem are removal of impulsive noise from images and clicks from audio signals. Two iterative algorithms were developed to minimize a proposed cost function.
We evaluated different aspects of our method through numerical experiments with comparison to other well-known methods.
We observed that our algorithm is fast and suitable for real-time applications and it outperforms the other methods in terms of the reconstruction quality and/or complexity. For images, the proposed algorithm works for any type of impulsive noise as opposed to other well-known methods that are suitable for either SPN or RVIN. For Future work, we can consider sparse noise in videos and block sparse noises such as block losses in the JPEG and MPEG images. Moreover, we wish to continue on this topic using machine learning techniques. We are currently working on soft thresholding that may improve the performance of recovery.

\appendices
\section{Proof of Lemma 1}
In order to compute this projection, we denote the concatenation of two matrices $\mathbf{U},\mathbf{V}$ by $[\mathbf{U},\mathbf{V}]$, and show the Frobenius inner product of two matrices, i.e., trace of their product, by $\langle\mathbf{U},\mathbf{V} \rangle$.

\begin{lemma}\label{lm:1}
If $\mathcal{D}$ is an orthogonal transformation, then $\langle[\mathcal{D}(\mathbf{U}),\mathbf{U}]  ,[\mathcal{D}(-\mathbf{V}),\mathbf{V}]\rangle=0$.
\end{lemma}

\begin{IEEEproof}
\begin{equation}\label{eq:7}
\begin{split}
   \langle[\mathcal{D}(\mathbf{U}),\mathbf{U}]  ,[\mathcal{D}(-\mathbf{V}),\mathbf{V}]  \rangle &= \langle\mathcal{D}(\mathbf{U}),\mathcal{D}(-\mathbf{V}) \rangle +
\langle\mathbf{U},\mathbf{V} \rangle \\
     & = \langle\mathbf{U},-\mathbf{V} \rangle + \langle\mathbf{U},\mathbf{V} \rangle = 0.
\end{split}
\end{equation}
\end{IEEEproof}

In order for $(\hat{\mathbf{X}},\hat{\mathbf{N}})$ to be the projection of $(\mathbf{X},\mathbf{N})$ onto the set $\mathrm{W}$, the error $(\mathbf{X},\mathbf{N}) - (\hat{\mathbf{X}},\hat{\mathbf{N}})$ has to be orthogonal to the set, i.e., it has to be orthogonal to the difference of any two members of $\mathrm{W}$.  The difference of any two members of $\mathrm{W}$ is $(\mathcal{D}(\mathbf{Y}-\mathbf{U_1}),\mathbf{U_1}) - (\mathcal{D}(\mathbf{Y}-\mathbf{U_2}),\mathbf{U_2}) = (\mathcal{D}(\mathbf{U_2}-\mathbf{U_1}),\mathbf{U_1}-\mathbf{U_2})$, which is of the form $(\mathcal{D}(-\mathbf{V}),\mathbf{V})$. By employing Lemma \ref{lm:1}, the error $(\mathbf{X},\mathbf{N}) - (\hat{\mathbf{X}},\hat{\mathbf{N}})$ has to be of the form $(\mathcal{D}(\mathbf{U}),\mathbf{U})$, where $\mathbf{U}$ is an auxiliary variable, so that the orthogonality is assured. Specifically:
\begin{equation}\label{eq:8}
\left\{ {\begin{array}{lr}
{\left( {{\mathbf{X}},{\mathbf{N}}} \right) - \left( {\hat{\mathbf{X}},\hat{\mathbf{N}}} \right) = \left( {{{\cal D}}\left( {\mathbf{U}} \right),\mathbf{U}} \right)}\\
{{{\cal D}^{ - 1}}\left( \hat{\mathbf{X}} \right) + \hat{\mathbf{N}} = \mathbf{Y}}
\end{array}} \right..
\end{equation}
The first equality holds as a result of Lemma \ref{lm:1}. The second equality also holds since $(\hat{\mathbf{X}},\hat{\mathbf{N}})\in\mathrm{W}$. One can easily omit the auxiliary variable $\mathbf{U}$
and obtain the desired result.

\section{Proof of Theorem 1}
Before we proceed with the proof, let $( {\mathbf{X^*}},{\mathbf{N^*}})$, with respective sparsity numbers of ($k_1,k_2$), be the minimizers of $f_{\lambda}$ for a given pair of $({{\mathbf{T}}_{\mathbf{1}}},{{\mathbf{T}}_{\mathbf{2}}})$. We will show that the cost function \eqref{eq:4} has greater values for binary matrices $({{\mathbf{T}}_{\mathbf{1}}},{{\mathbf{T}}_{\mathbf{2}}})$ other than $({\tilde{\mathbf{T}}_{\mathbf{1}}},{\tilde{\mathbf{T}}_{\mathbf{2}}})$. To this aim, we consider two cases:

Case 1:
\begin{equation}\label{eq:22}
\|(\mathbf{1}-{\mathbf{T}_{\mathbf{1}}})\odot{\mathbf{X^*}}\|_F = 0 \text{{ and }} \|(\mathbf{1}-{\mathbf{T}_{\mathbf{2}}})\odot{\mathbf{N^*}}\|_F = 0
\end{equation}

Case 2:
\begin{equation}\label{eq:24}
\|(\mathbf{1}-{\mathbf{T}_{\mathbf{1}}})\odot{\mathbf{X^*}}\|_F \neq 0 \text{{  or  }} \|(\mathbf{1}-{\mathbf{T}_{\mathbf{2}}})\odot{\mathbf{N^*}}\|_F \neq 0
\end{equation}

In the first case, we will show that $\|\text{vec}({\mathbf{T}_{\mathbf{1}}})\|_1 + \|\text{vec}({\mathbf{T}_{\mathbf{2}}})\|_1 \geq \tilde{k_1} + \tilde{k_2}$. Assume the opposite, that is:
\begin{equation}\label{eq:21}
\|\text{vec}({\mathbf{T}_{\mathbf{1}}})\|_1 + \|\text{vec}({\mathbf{T}_{\mathbf{2}}})\|_1  \le \tilde{k_1} + \tilde{k_2} -1,
\end{equation}
then $(\mathbf{1}-{\mathbf{T}_{\mathbf{1}}})$ and $(\mathbf{1}-{\mathbf{T}_{\mathbf{2}}})$ has at most $\tilde{k_1}+\tilde{k_2}-1$ zero entries in total. Therefore, in order for \eqref{eq:22} to be true, $\mathbf{X^*}$ and $\mathbf{N^*}$ can have at most $\tilde{k_1} + \tilde{k_2} -1$ non-zero entries in the corresponding zero elements of $(\mathbf{1}-{\mathbf{T}_{\mathbf{1}}})$ and $(\mathbf{1}-{\mathbf{T}_{\mathbf{2}}})$, respectively, in other words:
\begin{equation}\label{eq:23}
\begin{split}
   &\|\text{vec}({\mathbf{X^*}})\|_0 + \|\text{vec}({\mathbf{N^*}})\|_0 \leq \tilde{k_1} + \tilde{k_2} -1 \Rightarrow \\
   &{k_1} + {k_2} \leq \tilde{k_1} + \tilde{k_2} -1.
\end{split}
\end{equation}
Based on the assumption of the uniqueness of the sparsest solution, the sum of the sparsity numbers of every members of $\mathrm{W}$, except the sparsest one, is more than $\tilde{k_1}+\tilde{k_2}$; since $({\mathbf{X^*}},{\mathbf{N^*}})\in \mathrm{W}$, a contradiction is found. Consequently $\|\text{vec}({\mathbf{T}_{\mathbf{1}}})\|_1 + \|\text{vec}({\mathbf{T}_{\mathbf{1}}})\|_1  \geq \tilde{k_1} + \tilde{k_2}$, and the equality only holds for the sparsest member.

In the second case, since \eqref{eq:24} holds according to the definition $\varepsilon$ in \eqref{eq:13}, we have:
\begin{equation}\label{eq:25}
    \|(\mathbf{1}-{\mathbf{T}^i_{\mathbf{1}}})\odot{\mathbf{X^*}}^i\|_F^{2} + \|(\mathbf{1}-{\mathbf{T}^i_{\mathbf{2}}})\odot{\mathbf{N^*}}^i\|_F^{2}\geq \varepsilon >\lambda(\tilde{k_1} + \tilde{k_2}).
\end{equation}

One can easily see that in both cases, for all the members of $\mathrm{W}$, except for the sparsest one, ${f_{{\lambda}}}( {{\mathbf{X}},{\mathbf{N}},{\mathbf{T}_{\mathbf{1}}},{\mathbf{T}_{\mathbf{2}}}} )>
\lambda( \tilde{k_1} + \tilde{k_2}) ={f_{{\lambda }}}( \tilde{\mathbf{X}},\tilde{\mathbf{N}},{\tilde{\mathbf{T}}_{\mathbf{1}}},{\tilde{\mathbf{T}}_{\mathbf{2}}} )$. Hence the minimizer of $f_{{\lambda }}$ is indeed the sparsest member of $\mathrm{W}$.

\section{Proof of Theorem 2}
We use the following lemma in order to prove the theorem by contradiction.

\begin{lemma}\label{lm:2}
Let $\mathbf{\Phi}_\mathbf{1}$ and $\mathbf{\Phi}_\mathbf{2}$ be orthonormal bases for $\mathbb{R}^N$ and let
\begin{equation*}
  M(\mathbf{\Phi}_\mathbf{1},\mathbf{\Phi}_\mathbf{2}) = \text{\emph{sup}}\{|\langle \Phi_1,\Phi_2 \rangle|\: \Phi_1\in\mathbf{\Phi}_\mathbf{1} , \Phi_2\in\mathbf{\Phi}_\mathbf{2}\}.
\end{equation*}
Let $\Gamma_1$ be the set of indices of non-zero coefficients for $\mathbf{x}$ in basis 1, and $\Gamma_2$ be the set of indices of non-zero coefficients for $\mathbf{x}$ in basis 2. Then
\begin{equation}\label{eq:29}
  |\Gamma_1| + |\Gamma_2|\geq(1+M^{-1}).
\end{equation}
\end{lemma}

\begin{IEEEproof}
See Theorem 7.3 in \cite{donoho2001uncertainty}.
\end{IEEEproof}

Assume otherwise, that is, assume that there exists two members in $\mathrm{W}$ with sparsity numbers $(\tilde{k_1} , \tilde{k_2})$ and they are denoted by $(\mathbf{X}_\mathbf{1},\mathbf{N}_\mathbf{1})$ and $(\mathbf{X}_\mathbf{2},\mathbf{N}_\mathbf{2})$.
\begin{equation}\label{eq:15}
\begin{split}
   &{{\mathcal{D}}^{-1}}\left( \mathbf{X_1} \right) + \mathbf{N_1} = {{\mathcal{D}}^{-1}}\left( {\mathbf{X_2}} \right) + \mathbf{N_2} = \mathbf{Y} \Rightarrow \\
   &{{\mathcal{D}}^{-1}}\left( \mathbf{X_1}-{\mathbf{X_2}} \right) + (\mathbf{N_1}-\mathbf{N_2}) = \mathbf{0} \Rightarrow \\
   &\mathbf{A}\left( {\mathbf{X_1}-\mathbf{X_2}} \right)\mathbf{B} + (\mathbf{N_1}-\mathbf{N_2}) = \mathbf{0};
\end{split}
\end{equation}
by defining $\mathbf{X'} = {\mathbf{X_1}-\mathbf{X_2}}$ and $\mathbf{N'} = {\mathbf{N_1}-\mathbf{N_2}}$, with maximum sparsity numbers of $2\tilde{k_1}$ and $2\tilde{k_2}$, respectively, we will have:
\begin{equation}\label{eq:16}
\begin{split}
   &\mathbf{A}\,\mathbf{X'}\,\mathbf{B} + \mathbf{N'} = \mathbf{0} \Rightarrow \\
   &\text{vec}\left(\mathbf{A}\,\mathbf{X'}\,\mathbf{B}\right) + \text{vec}(\mathbf{N'}) =   \mathbf{0} \Rightarrow\\
   &\left(\mathbf{B}^\mathrm{T}\otimes \mathbf{A}\right)\text{vec}(\mathbf{X'}) + \text{vec}(\mathbf{N'}) = \mathbf{0};
\end{split}
\end{equation}
if $\mathbf{{x}} = \text{vec}(\mathbf{X'})$, $\mathbf{{n}} = \text{vec}(\mathbf{N'})$ and $\mathbf{C} = \mathbf{B}^\mathrm{T}\otimes \mathbf{A} \in \mathbb{R}^{mn\times mn}$,
we can conclude that:
\begin{equation}\label{eq:18}
{\mathbf{{x}} = -\mathbf{C}^{-1}\; \mathbf{{n}} = - \mathbf{C}^{\mathrm{T}}\; \mathbf{{n}}};
\end{equation}
in other words, $-\mathbf{n}$ is the representation of $\mathbf{x}$ in the orthonormal basis $\mathbf{C}$. According to Lemma 3, if $\mathbf{x}$ has $N_I$ and $N_C$ non-zero coefficients in orthonormal bases $\mathbf{I}$ (Identity matrix) and $\mathbf{C}$, respectively, then $N_I + N_C \geq (1+M(\mathbf{I},\mathbf{C})^{-1})$. Therefore:
\begin{equation}\label{eq:19}
\begin{split}
    &\|\mathbf{{x}}\|_0 + \|\mathbf{{n}}\|_0 =
    2\tilde{k_1} +2\tilde{k_2} \geq  \\
    &1+M(\mathbf{I},\mathbf{C})^{-1} =1+{\|\text{{vec}}(\mathbf{B}^\mathrm{T}\otimes \mathbf{A})\|_\infty}^{-1},
\end{split}
\end{equation}
which contradicts \eqref{eq:26}. Hence, ${\mathbf{x}} = \mathbf{0}$ and ${\mathbf{n}} = \mathbf{0}$. Since ${\mathbf{x}}$ and ${\mathbf{n}}$ are the vectorization of $\mathbf{X'}$ and $\mathbf{N'}$, respectively, $\mathbf{X'} = {\mathbf{X_1}-\mathbf{X_2}} = \mathbf{0}$ and $\mathbf{N'} = {\mathbf{N_1}-\mathbf{N_2}} = \mathbf{0}$. The uniqueness of the sparsest member of $\mathrm{W}$ is then concluded.

\bibliographystyle{IEEEtran}
\bibliography{IEEEabrv,bibi}

\end{document}